\documentclass[aip,reprint,twocolumn,nofootinbib]{revtex4-2}

\usepackage{amssymb}
\usepackage{amsfonts}
\usepackage{amstext}
\usepackage{graphicx}
\usepackage{amscd}
\usepackage{amsmath}
\usepackage{mathrsfs}
\usepackage{stmaryrd}
\usepackage{slashed}
\usepackage{wasysym}

\newcommand{\ihbar}{\imath \hbar}
\newcommand{\grad}{\overrightarrow{\mathrm{grad}}}
\newcommand{\rot}{\overrightarrow{\mathrm{curl}}}
\newcommand{\Div}{\mathrm{div}}

\newcommand{\llangle}{\langle \hspace{-0.2em} \langle}
\newcommand{\rrangle}{\rangle \hspace{-0.2em} \rangle}
\newcommand{\tr}{\mathrm{tr}}
\newcommand{\id}{\mathrm{id}}

\begin{document}

\title{Geometric phases, Everett's many-worlds interpretation of quantum mechanics, and wormholes}

\author{David Viennot}
\affiliation{Institut UTINAM (CNRS UMR 6213, Universit\'e de Franche-Comt\'e, Observatoire de Besan\c con), 41bis Avenue de l'Observatoire, BP 1615, 25010 Besan\c con cedex, France.}
\email{david.viennot@utinam.cnrs.fr}

\begin{abstract}
We present how the formalism of geometric phases in adiabatic quantum dynamics provides geometric realisations permitting to ``embody'' the Everett's many-worlds interpretation of quantum mechanics, including interferences between the worlds necessary for the probability changes and the decoherence processes needed to solve the preferred basis problem. We also show that this geometric realisation is intimately related to quantum gravity (especially to matrix models), showing that the many-world interpretation can be consistent with quantum gravity. The concept of wormhole borrowed from general relativity is central in this geometric realisation. It appears not only as an image by analogy to help the interpretations, but also as a true physical model of quantum wormhole in quantum gravity, the two ones being consistent which each other.
\end{abstract}

\maketitle

\section{Introduction}
Quantum mechanics is a successful theory to report experimental phenomena at the microscopic level and to predict results (or more precisely probabilities of results) through mathematical calculations. But it is well known that this theory is notably difficult to interpret. The reason for this is that our mental representations were forged by our sensory experiences at the macroscopic scale and are not useful for the microscopic world. It follows a large literature concerning the philosophy or the interpretation of quantum mechanics (see for example ref. \cite{Bitbol, Omnes, Espagnat}). Most quantum mechanics textbooks are written adopting the Copenhagen interpretation which consists to literally interpret the mathematical formalism. In the famous parable of the Schr\"odinger cat, when the cat state is $\frac{1}{\sqrt 2}(|\text{alive}\rangle+|\text{dead}\rangle)$, the Copenhagen interpretation states that the cat is ``simultaneously'' (in a superposition) of ``alive'' and ``dead'', the law of excluded middle being not applicable in quantum mechanics. But there are many other interpretations, as for example the many-worlds interpretation of Everett \cite{Everett} where the quantum state describes two worlds (or ``branches''), one where the cat is alive and the other where it is dead, which are compatible with the observation as long as the box enclosing the cat is closed. Due to the difficulty of interpretation, a lot of physicists prefer to adopt an attitude concerning the quantum mechanics within positivism, idealism (in the sense of Immanuel Kant) or instrumentalism.\\
In the present paper, we want to discuss the interest of the concept of geometric phases introduced by Berry \cite{Berry,Shapere} in the interpretation and the understanding of quantum dynamics from a realistic point of view. The interest of this subject is that it permits a geometric formulation of quantum dynamics, helping to ``visualise'' some abstract concepts. Moreover, as we will see in this paper, the concept of geometric phases is also relevant to discuss two other important problems in quantum mechanics, the quantum-classical transition by decoherence, and the problem of quantum gravity.
This paper is organised as follows. Section II presents the concept of geometric phases and the resulting geometrisation of quantum dynamics. Section III presents a generalisation of the concept of geometric phases (based on the adiabatic transport of several states) and its relation with the many-worlds interpretation of quantum mechanics. Section IV presents another generalisation (adapted to the problems of decoherence and entanglement), and discusses the question of the preferred basis of the many-worlds interpretation. Section V, we present a model of quantum gravity for which the concept of geometric phases is very important. Section VI discusses the question of the complexity of gauge theories through a comparison with geometric phases analysis and electrodynamics. Finally in a concluding section, we discuss the role of the geometric phases in a philosophical point of view concerning quantum mechanics. In the whole of this paper, we discuss the interest of the concept of ``wormhole between worlds'' in the interpretation of quantum dynamics. Section V presents also a physical model of quantum wormhole in quantum gravity (quantisation of wormholes issued from general relativity) which is in complete agreement with the interpretative concept of ``wormhole between worlds'', justifying the consistency between the two concepts.\\
The difficulties of interpretation of quantum mechanics induce difficulties to have intuitive mental images of quantum dynamics. From a pedagogical point of view, the geometric phases formalism provides geometric realisations of the many-worlds interpretation (by gauge fields similar to electromagnetic fields in ``parallel'' manifolds linked by ``wormholes'', see sections II to IV) which can help to built relevant mental images. Quantum gravity, which is one of the most important field of research of modern physics, is particularly difficult to discuss at an undergraduate level. Section V presents a simple model of quantum gravity easily related to usual quantum mechanics and general relativity via the geometric phase formalism. Moreover, the study of geometric phases can be used to introduce a discussion concerning epistemological questions about the ontology of quantum mechanics and electromagnetism as presented in section VI and in conclusion.

\section{Basic concepts about geometric phases}
\subsection{Berry-Simon geometric phases}
Following the principles of quantum mechanics, the states of a quantum system are described by an Hilbert space $\mathscr H$ (complex vector space), the observables modelling measurement devices are described by the set of selfadjoint linear operators (linear maps) of this Hilbert space $\mathcal L(\mathscr H)$, and the dynamics is described by the Schr\"odinger equation:
\begin{equation}
  \ihbar |\dot \psi\rangle = H(x(t)) |\psi(t)\rangle \qquad \psi \in \mathscr H
\end{equation}
where $H \in \mathcal L(\mathscr H)$ is the system Hamiltonian (energy observable). We suppose that $H$ is time-dependent through several classical variables $x \in \mathscr M$ belonging to a manifold $\mathscr M$ (we can view $\mathscr M$ as a domain of $\mathbb R^n$ for example). For example, in the Born-Oppenheimer approximation\cite{Moody, Teufel}, $\mathscr H$ is the state space of the electrons of a molecule whereas $x$ are the positions of the different nuclei (the movement of the nuclei being supposed correctly described by a classical evolution). Let $\{\lambda_a \in \mathbb R\}_a$ be the instantaneous eigenvalues of $H$ (supposed non-degenerate, continuous and differentiable with respect to $x$) and $\{|a,x\rangle \in \mathscr H\}_a$ be the associated normalised eigenvectors (supposed continuous and two times differentiable):
\begin{equation}
  H(x)|a,x\rangle = \lambda_a(x)|a,x\rangle
\end{equation}
In the strict adiabatic approximation, if $|\psi(0)\rangle = |a,x(0)\rangle$ we have\cite{Teufel}
\begin{equation}
  |\psi(t) \rangle \simeq e^{-\ihbar^{-1} \int_0^t \lambda_a(x(t'))dt'} e^{-\imath \int_0^t A_i(x(t')) \dot x^i(t') dt'} |a,x(t) \rangle
\end{equation}
(we adopt the Einstein convention, the repeated upper/lower index $i$ is equivalent to a summation) where $A_i = -\imath \langle a,x|\partial_i|a,x \rangle \in \mathbb R$ ($\partial_i \equiv \frac{\partial}{\partial x^i}$). The approximation being valid if:
\begin{equation}\label{adiabcond}
  \sup_t \max_{b\not=a} \left|\frac{\hbar \langle b,x|\partial_i|a,x\rangle \dot x^i}{\lambda_b-\lambda_a}\right| \ll 1
\end{equation}
or equivalently if $\tau \gg T$ where $\tau = \inf_t \min_{b \not=a} (\lambda_b-\lambda_a)/\hbar$ (for $b$ such that the non-adiabatic couplings $\langle b,x|\partial_i|a,x\rangle$ are not all zero) is the smallest Rabbi period of the quantum system (characteristic time of the quantum evolution) and where $T$ is the duration of the interaction (characteristic time of the classical evolution $t \mapsto x(t)$). The adiabatic approximation is then valid if the classical evolution $t \mapsto x(t)$ is slow. The adiabatic approximation is equivalent to state that if the evolution of $x$ is slow, the quantum system stays in the same instantaneous ``equilibrium'' state $|a,x(t)\rangle$ (as in classical mechanics, a ball on a plate stays at the same place on the plate if this one is moved very slowly). $e^{-\ihbar^{-1} \int_0^t \lambda_a(x(t'))dt'}$ is called dynamical phase whereas $e^{-\imath \int_0^t A_i(x(t')) \dot x^i(t') dt'}$ is called geometric phase\cite{Simon} because it depends only on the shape of the path $\mathscr C \subset \mathscr M$ drawn by $t\mapsto x(t)$: $\int_0^t A_i(x(t')) \dot x^i(t') dt' = \int_{\mathscr C} \vec A \cdot d\vec x$, with $\vec A = -\imath \langle a,x|\vec \nabla|a,x\rangle$ viewed as a (tangent) vector on $\mathscr M$.
\subsection{Magnetic analogy}
Since the phase of the eigenvector is arbitrary, we can consider phase changes as
\begin{equation}
  |a,x\rangle' = e^{\imath \chi(x)} |a,x\rangle
\end{equation}
and then
\begin{equation}
  \vec A' = \vec A + \grad \chi
\end{equation}
This formula is similar to the gauge change formula for a magnetic potential $\vec A$. By analogy we can then introduce a magnetic field $\vec F = \rot \vec A$ (invariant under gauge/phase change). The adiabatic phase phenomenon is then similar to the Aharonov-Bohm effect\cite{Aharonov} associated with the transport of a charged particle in a magnetic potential. By using the closure relation associated with the eigenvector basis, we can prove that the components of $\vec F$ are
\begin{eqnarray}
  F_{k} {\varepsilon^k}_{ij} & = & \partial_i A_j - \partial_j A_i \\
  & = & -\imath (\partial_i\langle a,x|)(\partial_j|a,x\rangle) + c.c. \\
  & = & -\imath \sum_{b \not=a} \overline{\langle b,x|\partial_i|a,x\rangle} \langle b,x|\partial_j|a,x\rangle + c.c.
\end{eqnarray}
($\varepsilon_{kij}$ being the Levi-Civita symbol). $\vec F$ is then the field of the non-adiabatic couplings. Note that $H|a,x\rangle=\lambda_a|a,x\rangle \Rightarrow (\partial_iH)|a,x\rangle + H\partial_i|a,x\rangle = (\partial_i \lambda_a)|a,x\rangle + \lambda_a \partial_i|a,\rangle$, and by projecting this equation onto $\langle b,x|$ we have
\begin{equation}
  \langle b,x|\partial_i|a,x\rangle = \frac{\langle b,x|\frac{\partial H}{\partial x^i}|a,x\rangle}{\lambda_a(x)-\lambda_b(x)}
\end{equation}
If at a point $x_*$ another eigenvalue crosses $\lambda_a$, e.g. $\lambda_a(x_*)=\lambda_b(x_*)$ ($b\not=a$), then $\vec F$ diverges at $x_*$. The crossing point $x_*$ appears then as a magnetic monopole\cite{Moody} and to satisfy the condition eq.(\ref{adiabcond}) the path $\mathscr C$ must avoid the neighbourhood of $x_*$. 

\subsection{Generalisations of geometric phases}
We can also consider a degenerate eigenvalue $\lambda_a$. In that case, $\vec A$ becomes matrix valued \cite{Wilczek} and is analogous to a non-abelian gauge field (for a degenerescence degree equal to 2, $\vec A$ is analogous to a weak nuclear interaction potential (in the Yang-Mills model), and for a degree equal to 3, $\vec A$ is analogous to a strong nuclear interaction potential).\\
We can also consider non-adiabatic but cyclic dynamics\cite{Aharonov2}  $\ihbar |\dot \psi \rangle = H(t)|\psi(t) \rangle$ with $H(T)=H(0)$. Let $n=\dim \mathscr H$ be the dimension of the Hilbert space and $\{|i\rangle\}_{i=0,...,n-1}$ be a basis of $\mathscr H$. We have $|\psi \rangle = \sum_{i=0}^{n-1} w^i|i\rangle$ with $w^i \in \mathbb C$. Normalised quantum states can be then represented by $n-1$ complex numbers $z^i = \frac{w^i}{w^0}$. The set of all possible configurations of these numbers generates a (complex) manifold $\mathscr M$ (which replaces the manifold of the classical parameters of the adiabatic approximation). For example with a two level system, we have $|\psi\rangle = w^0|0\rangle+w^1|1\rangle$ with $|w^0|^2+|w^1|^2=1$. We can then choose $w^0 = \cos\frac{\theta}{2}$ and $w^1 = e^{\imath \varphi} \sin \frac{\theta}{2}$ and then $z^1 =  e^{\imath \varphi} \tan \frac{\theta}{2}$. $\mathscr M$ is then a sphere (the Bloch sphere), $(\theta,\varphi)$ being its spherical coordinates system whereas $(\Re\mathrm{e}(z^1),\Im\mathrm{m}(z^1))$ are the coordinates in the plane of the stereographic projection of the sphere. $|\tilde \psi(t)\rangle = e^{-\imath \int_0^t \langle \psi|H|\psi\rangle dt'} e^{-\imath \int_0^t (A_i \dot z^i + A_{\bar i} \dot{\bar z}^i) dt'} |\psi(t)\rangle$ is the cyclic state associated with the cyclic evolution ($H(T)=H(0) \Rightarrow |\tilde \psi(T)\rangle = |\tilde \psi(0)\rangle$), with $A_i = -\imath\langle \psi|\frac{\partial}{\partial z^i}|\psi \rangle$ and $A_{\bar i} = -\imath\langle \psi|\frac{\partial}{\partial \bar z^i}|\psi \rangle$. So we can consider non-adiabatic dynamics in the same manner than adiabatic dynamics, by just replacing the manifold of the classical parameters by the (complex) manifold of the normalised quantum states. But since this one is complicate to describe (except for the two level system case), we prefer for the sake of simplicity to consider only the adiabatic case in the sequel of this paper.

\section{Adiabatic transport of several eigenstates}
Consider a set of $N < \dim \mathscr H$ instantaneous eigenvalues $\{\lambda_a\}_{a \in\{1,...,N\}}$  (for convenience these ones are labelled from $1$ to $N$ but they are not necessarily the smallest eigenenergies of the quantum system). A generalisation of the adiabatic theorem\cite{Nenciu} states that if $|\psi(0)\rangle = \sum_{a=1}^N c_a|a,x(0)\rangle$, then we have the following adiabatic approximation:
\begin{equation}\label{nonadtransp}
  |\psi(t)\rangle \simeq \sum_{a,b=1}^{N} U_{ba}(t) c_a |b,x(t)\rangle
\end{equation}
where $U$ is $N \times N$ unitary matrix (a ``non-abelian phase'') such that
\begin{equation}
  \imath \frac{dU(t)}{dt} = (-\hbar^{-1} E(x(t))+A_i(x(t)) \dot x^i(t)) U(t)
\end{equation}
with $E=\mathrm{diag}(\lambda_1,...,\lambda_N)$ is the diagonal matrix of the eigenvalues and $A_i$ is the matrix of elements $[A_i]_{ab} = \langle a,x|\partial_i|b,x\rangle$. This adiabatic approximation is valid if 
\begin{equation}
  \sup_t \max_{a\in\{1,...,N\},b>N} \left|\frac{\hbar \langle b,x|\partial_i|a,x\rangle \dot x^i}{\lambda_b-\lambda_a}\right| \ll 1
\end{equation}
In general $[E,A]\not=0$ and it cannot be possible to separate the non-abelian dynamical and geometric phases, generating a complicated geometric description\cite{Viennot1}. Here we want to consider a special case where $N=2$ and where the path $\mathscr C \subset \mathscr M$ does not meet any eigenvalue crossing except at time $t_*$ where $\lambda_1(x(t_*)) = \lambda_2(x(t_*))$ (supposed to be a conical crossing). Suppose now that the condition eq.(\ref{adiabcond}) is satisfied for any time $t$ except in a small neighbourhood of $t_*$ and that $\mathscr C$ is differentiable in the neighbourhood of $t_*$. In these conditions we can prove\cite{Teufel} that eq.(\ref{nonadtransp}) is reduced to
\begin{equation}\label{adiabtransp2}
  |\psi(t) \rangle \simeq  c_1 e^{-\imath \varphi_1(0,t)} |1,x(t)\rangle + c_2 e^{-\imath \varphi_2(0,t)} |2,x(t)\rangle
\end{equation}
$\forall t<t_*$ and
\begin{eqnarray}
  |\psi(t) \rangle & \simeq & c_1 e^{-\imath (\varphi_1(0,t_*)+\varphi_2(t_*,t))} |2,x(t)\rangle \nonumber \\
  & & \quad + c_2 e^{-\imath (\varphi_2(0,t_*)+\varphi_1(t_*,t))} |1,x(t)\rangle
\end{eqnarray}
$\forall t>t_*$; with $\varphi_a(s,t) = \hbar^{-1} \int_s^t \lambda_a(x(t'))dt' + \int_s^t A_i^a(x(t'))\dot x^i(t') dt'$, $\vec A^a = -\imath \langle a,x|\vec \nabla|a,x\rangle$. In the electromagnetic analogy of the adiabatic transport, everything happens as if the manifold $\mathscr M$ were endowed with a structure of two ``worlds'' $\{\mathscr M^a\}_a$ in which we find two different magnetic fields $\vec F^a = \rot \vec A^a$. While we do not measure the energy of the quantum system, permitting to the state to be a superposition of $|1,\vec x\rangle$ and $|2,\vec x\rangle$ (with probabilities $|c_1|^2$ and $|c_2|^2$), $\{(\mathscr M^a, \vec A^a)\}_a$ can be viewed as a manifestation of the two worlds of the Everett interpretation. There is a world $\mathscr M^1$ where the charged particle (equivalent to the quantum system in the magnetic analogy) sees a magnetic field $\vec F^1$ and another world $\mathscr M^2$ where it sees a magnetic field $\vec F^2$. The magnetic monopole at $x_* = x(t_*)$ is then the manifestation of the interferences between the two worlds, since the ``part'' of the particle being initially in the world $\mathscr M^1$ goes in $\mathscr M^2$ when the path $\mathscr C$ passes through the monopole (and reciprocally for the ``part'' being initially in the world $\mathscr M^2$). From a geometric point of view, $\mathscr M^1$ and $\mathscr M^2$ can be viewed as two parallel spaces linked by a kind of ``wormhole'' at $x_*$. Note that in general relativity, since a wormhole appears for an observer as a source of magnetic field (the magnetic lines entering in the wormhole go out in another sheet of spacetime), it appears as a magnetic monopole\cite{Frankel}. Note that the passage from a world to another one (the inversion of the occupation probabilities) does not need that $\mathscr C$ passes exactly through $x_*$, it needs only that $\mathscr C$ passes through a small neighbourhood of $x_*$, the transition being then ensured by a Landau-Zener transition\cite{Teufel}. The values of the occupation probabilities are sustained in this adiabatic approximation, but if we suppose that $\mathscr C$ is not differentiable at $x_*$ (supposed to be a conical crossing), then for $t>t_*$ we have\cite{Boscain}
\begin{eqnarray}
  |\psi(t) \rangle & \simeq & c_1' e^{-\imath (\varphi_1(0,t_*)+\varphi_2(t_*,t))} |2,x(t)\rangle \nonumber \\
  & & \quad + c_2' e^{-\imath (\varphi_2(0,t_*)+\varphi_1(t_*,t))} |1,x(t)\rangle
\end{eqnarray}
with $c_1' = \cos\alpha\, c_2 + e^{\imath \beta} \sin \alpha\, c_1$ and $c_2' = \cos\alpha\, c_1 + e^{-\imath \beta} \sin \alpha\, c_2$, where $2\alpha$ is the angle between the tangent vectors of $\mathscr C$ at $x_*$ for $t\to t_*^-$ and for $t\to t_*^+$ and $\beta = \arg\langle 1,x|\partial_i|2,x\rangle_{|t\to t_*^-} \dot x^i_{|t \to t_*^+} + \pi$. The interferences between the two worlds can then modify the occupation probability values. With a higher speed $\dot{\vec x}$, in place of an instantaneous transition at $t_*$ we have a smooth transition in a larger neighbourhood of $t_*$. In other words, the size of the region around the magnetic monopole inducing a transition between the worlds increases with the particle speed. The general formula eq.(\ref{nonadtransp}) corresponds to this case where the ``wormhole influence'' is extended on the whole of $\mathscr C$. The discussion can be easily generalised with $N>2$ and with several crossings (monopoles, wormholes).\\

\begin{figure}
  \includegraphics[width=7cm]{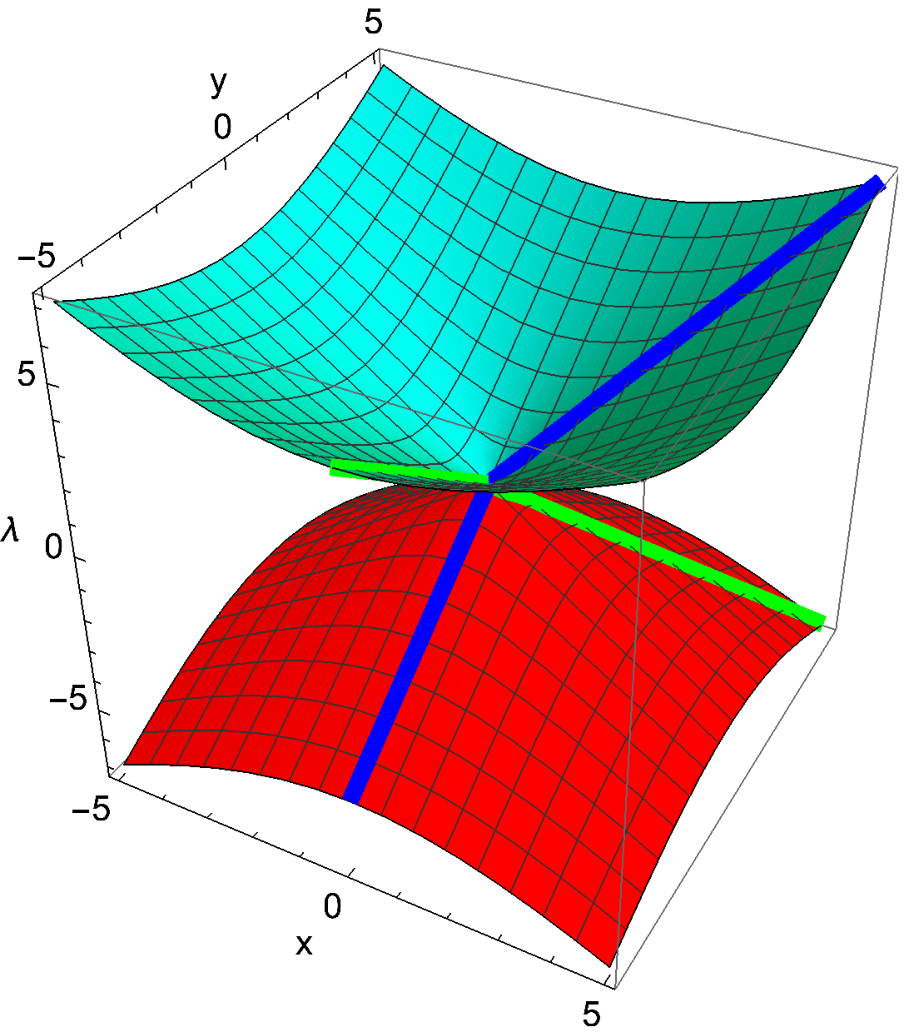}\\
  \includegraphics[width=7cm]{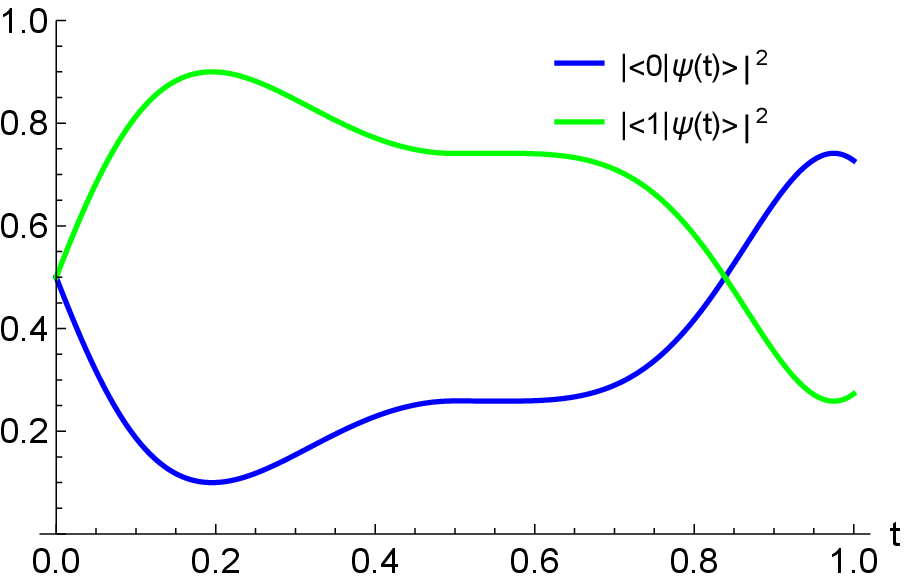}\\
  \includegraphics[width=7cm]{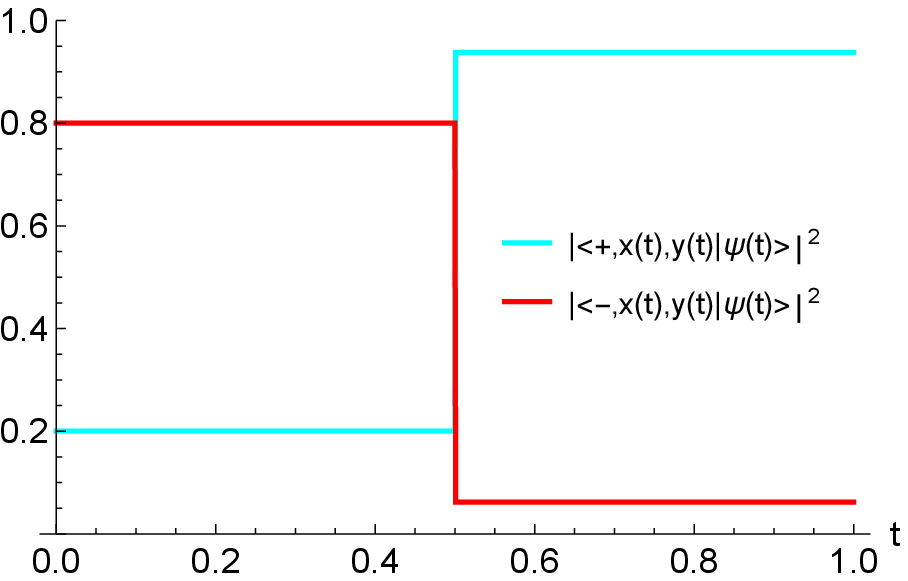}\\
  \caption{\label{adiabsurfaces} Up: $\mathscr M^\pm$ embedded in $\mathbb R^3$ and the representation of a path $\mathscr C$ passing through the ``wormhole'' (magnetic monopole) with a bend (blue for the part of the particle starting on $\mathscr M^-$ and green for the part starting on $\mathscr M^+$). Middle: occupation probabilities in the bare basis $(|0\rangle,|1\rangle)$ in which the Hamiltonian eq.(\ref{Hexpl}) is written, for the path $\mathscr C$ and the initial condition $|\psi(0)\rangle = \sqrt{0.2}|+,x(0),y(0)\rangle + \sqrt{0.8}|-,x(0),y(0)\rangle$. Down: occupation probabilities in the instantaneous eigenbasis $(|\pm,x(t),y(t)\rangle)$. The three plots are drown in atomic units. In the bare picture, the interpretation of the evolution of the quantum probabilities are not clear. In the adiabatic picture, we can explain the evolution by the geometric representation (from $t=0$ to $t=0.5\ \mathrm{a.u.}$ the particle is in the superposition of two worlds $\mathscr M^\pm$ with the initial probabilities, at $t=0.5\ \mathrm{a.u.}$ the passage by the wormhole induces the exchange of the worlds and the bend of the curve mixes the probabilities).}
\end{figure}

The adiabatic picture provides then a geometric realisation of the Everett interpretation of the quantum mechanics, available in quantum dynamics. The many worlds are parallel manifolds $\{\mathscr M^a\}$ where live different magnetic potentials $\vec A^a$. The interferences between the worlds are localised at the magnetic monopoles (common at two worlds) which play the role of wormholes between two worlds. We can illustrate this by the simplest example of the following Hamiltonian:
\begin{equation}\label{Hexpl}
  H(x,y) = \left(\begin{array}{cc} 0 & x-\imath y \\ x + \imath y & 0 \end{array} \right)
\end{equation}
of eigenvalues and associated eigenvectors:
\begin{eqnarray}
  \lambda_{\pm}(x,y) & = & \pm r \qquad  \\
  |+,x,y\rangle & = & \frac{1}{\sqrt 2 r} \left(\begin{array}{c} r \\ x+\imath y \end{array} \right) = \frac{1}{\sqrt 2} \left(\begin{array}{c} 1 \\ e^{\imath \theta} \end{array} \right) \\
  |-,x,y\rangle & = & \frac{1}{\sqrt 2 r} \left(\begin{array}{c} x-\imath y \\ -r \end{array} \right) = \frac{1}{\sqrt 2} \left(\begin{array}{c} e^{-\imath \theta} \\ -1 \end{array} \right)
\end{eqnarray}
with $x+\imath y = re^{\imath \theta}$. The magnetic potentials are
\begin{equation}
  \vec A^\pm = \mp \frac{y \vec e_x - x \vec e_y}{2r^2} = \pm \frac{1}{2r} \vec e_\theta
\end{equation}
We can remark that in this case $\vec F^\pm = \rot \vec A^\pm = \vec 0$ (except at $0$), but $\oint_{\mathscr C} \vec A^\pm \cdot d\vec x =  \pm \frac{1}{2}$ for any closed curve $\mathscr C$ surrounding $r=0$ one time in the counterclockwise direction. Due to the magnetic monopole of magnetic charge $\frac{1}{2}$ at $r=0$ (where $\vec A^\pm$ is singular), even if the magnetic field is zero, the circulation of the magnetic potential is not zero, as in the Aharonov-Bohm effect\cite{Aharonov}. A useful representation of the worlds $\{\mathscr M^\pm\}$ consists to use their embedding in $\mathbb R^3$ provided by $(x,y) \mapsto (x,y,\lambda_\pm(x,y))$. The wormhole is then a crossing point of the two surfaces, see fig.\ref{adiabsurfaces}
The adiabatic geometric picture is not in contradiction with the fact that the worlds in the Everett's interpretation are emerging structures and not parallel ``spacetimes'' (the system, the observer, the spacetime, the universe remain single but presenting both a relative state as a superposition of branches in the language used in the Everett's original article\cite{Everett}). Firstly, we can note that the path $\mathscr C$ and the time with which it is traveled are common to the two worlds (same path projected on $\mathscr M^\pm$ in fig.\ref{adiabsurfaces}). The ``spacetime'' is then common as in the usual Everett's interpretation. Moreover, the structure $(\{\mathscr M^\pm, A^\pm\})$ emerges from the diagonalisation of $H$ (the potential measures realised by the observer concern the instantaneous energies). In his original paper, Everett presents his reasoning leading to his many worlds interpretation by introducing explicit relative states as bipartite states of the system and the observer. This entanglement is present in the adiabatic picture by the labelling of the quantum eigenstates by the classical parameters $x$ (``state'' of the observer in the meaning of the ``state'' of the experimental device used by the observer to control the quantum system). It is possible to use a different formalism (not directly related to the many-world formulation) permitting to built states associated with the geometric phase phenomenon which present explicitly entanglement between the system and the observer and are consistent with the electromagnetic analogy (see appendix \ref{RSF}).\\

In the context of the adiabatic approximation, the preferred basis problem takes a special form. This problem is the fact that the worlds of the Everett's interpretation are associated with a particular basis. But from the point of view of linear algebra, the mathematical underlying structure of quantum mechanics, the basis choice is arbitrary. In the context of the adiabatic assumptions, the many-worlds $\{\mathscr M^a , A^a\}$ (as manifolds endowed with a magnetic field) are determined by the instantaneous eigenstates $\{|a,x\rangle\}$. These states permit the geometric formulation of the dynamics as a particle moving in many-worlds and submitted to magnetic fields and ``wormholes'' between the worlds. Why this choice of basis is preferred? Suppose that the Hilbert space $\mathscr H$ is infinite dimensional (this is the current true situation without any approximation). We can decide to choose any basis to represent the system. For example, we choose the eigenbasis at time $t=0$. By a change of basis in eq.(\ref{adiabtransp2}) we have:
\begin{equation}
  |\psi(t)\rangle \simeq \sum_{a=1}^{\infty} K_a(t)|a,x(0)\rangle
\end{equation}
with
\begin{eqnarray}
  K_a(t) & = & c_1e^{-\imath \varphi_1(0,t)}\langle a,x(0)|1,x(t)\rangle \nonumber \\
  & & + c_2e^{-\imath \varphi_2(0,t)}\langle a,x(0)|2,x(t)\rangle
\end{eqnarray}
Due to the evolution $t \mapsto x(t)$, if initially only two states (in the basis $(|a,x(0)\rangle)_a$) are populated, a lot of them are populated during the dynamics. The preferred basis, the $x$-dependent adiabatic basis $(|a,x\rangle)_a$, is the basis involving the use of the minimal number of states relative to the observer in order to represent the dynamics (``relative'' in the meaning of being associated with a possible measurement, the choose of a time-dependent basis including $|\psi(t)\rangle$ is not relative to the observer since the state of the quantum system is not necessarily associated with a possible measurement by the observer and it depends on the history of the quantum system). This is an import point of the many-worlds interpretation, the different worlds are the manifestation of the entanglement between the observed quantum system and the observer. $|\psi\rangle$ is a relative state representing the correlation between the system and the observer. In the adiabatic formalism, this is done by the dependency of the basis on $x$. These classical parameters are measured quantities by the observer (as the nuclei position in the Born-Oppenheimer approximation for example) or are control parameters modulated by the observer (as the intensity, the frequency, the phase or the polarisation direction of an electric field in laser control of an atom for example). Moreover, in a pragmatic point of view, we can consider the preferred basis as being the one in which the quantum dynamics is intelligible (as in the example presented fig. (\ref{adiabsurfaces})). One might think that this pragmatic attitude is unsatisfactory since we would expect that an interpretation concerns the ``reality of the Nature''. But if we consider that the many-worlds interpretation is at first a representation of relative states between the quantum system and the observer, it is natural to consider that the preferred basis be intelligible for the observer.\\
An argument\cite{Wallas} states that the solving of the preferred basis problem requires the incorporation of the decoherence. In next section, we explore the incorporation of this one in the framework of geometric phases.

\section{Geometric phases of open quantum systems}
\subsection{Weak adiabatic transport}
In the reality the quantum systems are not isolated and are submitted to interaction with a large environment composed by the matter, the electromagnetic field and the vacuum fluctuations surrounding the system. This environment can be modelled by a large quantum system described by an Hilbert space $\mathscr E$. The bipartite system, composed by the observed system and its environment, is then described by the Hilbert space $\mathscr H \otimes \mathscr E$ with an Hamiltonian:
\begin{equation}
  H_{tot}(x) = H(x) \otimes 1_{\mathscr E} + 1_{\mathscr H} \otimes H_{\mathscr E}(x) + \epsilon V(x)
\end{equation}
where $H$ is the Hamiltonian of the observed system, $H_{\mathscr E}$ is the Hamiltonian of the environment and $\epsilon V$ is the interaction operator between the system and the environment. Note that we have considered the possibility that the classical control of the observer $x$ acts also on the environment and on the interaction. The state $|\psi \rrangle \in \mathscr H \otimes \mathscr E$ of the bipartite system obeys to the Schr\"odinger equation:
\begin{equation}
  \ihbar |\dot \psi \rrangle = H_{tot}(x(t))|\psi(t)\rrangle
\end{equation}
But the observer observes and measures only the system, not the environment. Moreover $\mathscr E$ is large (large number of quantum degrees of freedom) and impossible to know. So the information concerning the state of the environment is unknown to the observer. The relative state between the system and the observer is then given by the partial trace over $\mathscr E$ of the bipartite system state:
\begin{eqnarray}
  \rho(t) & = & \tr_{\mathscr E} |\psi(t)\rrangle \llangle \psi(t)| \\
  & \equiv & \sum_{a,b} \sum_{\alpha=1}^{\infty} \llangle a,\alpha |\psi(t)\rrangle \llangle \psi(t)| b,\alpha\rrangle |a\rangle\langle b|
\end{eqnarray}
$(|\alpha\rangle)$ is a basis of $\mathscr E$ and $(|a\rangle)$ a basis of $\mathscr H$ (the partial trace is independent of the choice of these bases), the latin labels being associated with the system and the greek labels with the environment. The partial trace erases the information concerning the state of the environment. The mixed state $\rho$ represents a statistical mixture of quantum states. For example, with a 2-level system, the density matrix is $\rho = \left(\begin{array}{cc} p_0 & \bar c \\ c & p_1 \end{array} \right)$ in a basis $(|0\rangle,|1\rangle)$, in which $p_a$ is the probability of occupation of the state $|a\rangle$ and $|c|$ is the coherence. If $|c|$ is large the state $\rho$ is close to a quantum superposition of the states $(|0\rangle,|1\rangle)$ (the state is $|0\rangle$ \textbf{and} $|1\rangle$) and its purity $\mathcal P(\rho) = \tr(\rho^2)$ is close to 1; whereas if $c=0$ the state $\rho$ is a statistical mixture of the states (as in classical statistical physics, the state is $|0\rangle$ \textbf{or} $|1\rangle$, it is unknown due to the lack of information concerning $\mathscr E$) and its purity is small (except if $p_0$ or $p_1$ is close to 0). Due to the interaction $V$, during the dynamics the state of the bipartite system becomes entangled and by consequence the purity and the coherence of $\rho$ decrease (a phenomenon called decoherence\cite{Breuer}). The von Neumann entropy $S(\rho) = - \tr(\rho \ln \rho)$ measures the degree of entanglement of the bipartite state and the lack of information of the observer concerning the state of the system due to the lack of knowledge on the environment.\\
In the perturbative regime ($\epsilon \ll 1$) a weak adiabatic theorem\cite{Viennot2} takes into account the effects of the environment. Let $(|a,\alpha,x\rrangle_\epsilon)$ be the instantaneous eigenvectors of $H_{tot}$, with $|a,\alpha,x\rrangle_\epsilon = |a,x\rangle \otimes |\alpha,x\rangle + \mathcal O(\epsilon)$ where $(|a,x\rangle)$ are the eigenvectors of $H(x)$ (states of the system) and $(|\alpha,x\rangle)$ are the eigenvectors of $H_{\mathscr E}(x)$ (states of the environment). The associated eigenvalues are denoted by $(\lambda_{a\alpha}^{\epsilon}(x))$, $(\mu_a(x))$ and $(\nu_\alpha(x))$. Under an assumption similar to eq.(\ref{adiabcond}) for $(\nu_\alpha)$ (the time characterising the classical evolution $t\mapsto x(t)$ is large with respect to the time characterising the dynamics of the environment, meaning here that the environment is sufficiently large for the external actions cannot induce on it sudden changes) and under the assumption $|\mu_b+\nu_\beta-\mu_c-\nu_\alpha|\gg \epsilon$ $\forall c, \forall  (b\beta)\not=(c\alpha)$ (no quasi-resonance between the transitions of the system and the ones of the environment involving the state  $|\alpha,x\rangle)$ then if $|\psi(0)\rrangle = |a,\alpha,x(0)\rrangle_\epsilon$ we have at the first order of perturbation:\cite{Viennot2}
\begin{equation}
  \rho(t) \simeq U_E(t) U_A(t) \rho_{a\alpha}^\epsilon(x(t)) U_A^\dagger(t) U_E^\dagger(t)
\end{equation}
where $\rho_{a\alpha}^\epsilon(x) = \tr_{\mathcal E} |a,\alpha,x\rrangle_{\epsilon \epsilon}\llangle a,\alpha,x|$ is the mixed state associated with the initial bipartite eigenstate, and where the operators $U_E$ and $U_A$ are defined by
\begin{eqnarray}
  \ihbar \dot U_E & = & E^\epsilon(x(t)) U_E(t) \\
  \imath \dot U_A & = & U_A(t) \mathcal A_i(x(t)) \dot x^i(t)
\end{eqnarray}
with
\begin{equation}
  E^\epsilon(x) = \sum_b \lambda^\epsilon_{b\alpha} |b,x\rangle \langle b,x| + \mathcal O(\epsilon)
\end{equation}
and
\begin{equation}
  \vec {\mathcal A}(x) \rho_{a\alpha}^{\epsilon}(x) = -\imath \tr_{\mathscr E}\left(P_{\bullet \alpha}(x)\vec \nabla |a,\alpha,x\rrangle_{\epsilon \epsilon}\llangle a,\alpha, x|\right)
\end{equation}
$P_{\bullet \alpha}(x) = \sum_b |b,\alpha,x\rrangle_{\epsilon \epsilon}\llangle b,\alpha,x|$. The geometric phase $U_A$ and the potential $\vec {\mathcal A}$ are  operators of the system in this weak adiabatic transport. We find the usual magnetic potential $\vec A = -\imath {_\epsilon}\llangle a,\alpha,x|\vec \nabla|a,\alpha,x\rrangle_\epsilon = \tr(\rho_{a\alpha}^\epsilon(x) \vec {\mathcal A}(x))$ as the statistical averaging of $\vec {\mathcal A}$. Let $(|b_{a\alpha},x\rangle_\epsilon)$ be the diagonalisation basis of $\rho_{a\alpha}^\epsilon(x)$ ($\rho_{a\alpha}^\epsilon(x) = \sum_bp_b^{a\alpha}|b_{a\alpha},x\rangle_{\epsilon\epsilon}\langle b_{a\alpha},x|$), we have then
\begin{equation}
  \vec A = \sum_b p_b^{a\alpha} \vec {\mathcal A}_{bb}
\end{equation}
with $\vec {\mathcal A}_{bb} = {_\epsilon}\langle b_{a\alpha},x|\vec {\mathcal A}|b_{a\alpha},x\rangle_\epsilon$. Due to the lack of information $S(\rho_{a\alpha}^\epsilon) = - \sum_b p_b^{a\alpha} \ln p_b^{a\alpha}$, the particle is submitted to the magnetic potentials $\vec {\mathcal A}_{bb}$ with probabilities $p_b^{a\alpha}$. $\vec {\mathcal A}$ can be then viewed as a random variable of magnetic potential, $\vec A$ being its mean value.\\

We return to the question of the preferred basis of the Everett's interpretation. In the weak adiabatic transport the dynamics is described with the eigen mixed state $\rho_{a\alpha}^\epsilon$ (by conjugation with the operator valued dynamical and geometric phases). Following the argument concerning the preferred basis\cite{Wallas}, this one can be defined as being the one in which no coherence appears (no quantum superposition). This is then the diagonalisation basis of $\rho_{a\alpha}^\epsilon(x)$. But
\begin{equation}
  \lim_{\epsilon \to 0} \rho_{a\alpha}^\epsilon(x) = |a,x\rangle \langle a,x|
\end{equation}
Without the effect of the environment, the preferred basis is well the eigenbasis of $H(x)$ as in the previous sections. Note that due to the operator valued dynamical and geometric phases, $\rho(t)$ is not diagonal in the basis $(|b_{a\alpha},x\rangle_\epsilon)$. But $\rho(t)$ depends on the followed path $t\mapsto x(t)$, it cannot define a preferred basis independent of the history of the system. If the environment induces a decoherence process, the external action represented by the evolution of $x$ can recreate some coherence (via the action of the operator valued phases). 

\subsection{Adiabatic fields}
As for the geometric phase associated with closed systems, it is possible to generalise the operator valued geometric phase of open quantum systems to non-adiabatic evolutions\cite{Viennot3}. $\mathscr M$ is then replaced by the manifold of the density matrices and the potential can be defined directly with the density matrices as being  solution of $-\imath \vec \nabla \rho = \vec {\mathfrak A} \rho + \rho \vec {\mathfrak A}^\dagger$. In the adiabatic case, this potential is
\begin{equation}
  \vec {\mathfrak A}(x) \rho_{a\alpha}^\epsilon(x) = -\imath \tr_{\mathscr E}\left(\vec \nabla|a,\alpha,x\rrangle_{\epsilon \epsilon} \llangle a,\alpha,x|\right)
\end{equation}
$\vec {\mathcal A}$ and $\vec {\mathfrak A}$ are not exactly the same operator, even if $\tr(\rho_{a\alpha}^\epsilon \vec {\mathfrak A}) = \tr(\rho_{a\alpha}^\epsilon \vec {\mathcal A}) = \vec A$, but in the assumption of the weak adiabatic approximation we have\cite{Viennot4}:
\begin{equation}
  \mathfrak A_i \dot x^i \simeq \mathcal A_i \dot x^i
\end{equation}
The two potentials define two operator valued adiabatic fields:
\begin{eqnarray*}
  \vec {\mathcal B} & = & \rot \vec {\mathfrak A} +\imath \vec {\mathfrak A} \wedge \vec {\mathfrak A} \\
  \vec {\mathcal F} & = & \rot \vec {\mathcal A} +\imath \vec {\mathcal A} \wedge \vec {\mathcal A} - \vec {\mathcal B}
\end{eqnarray*}
Since the potentials are operator valued with components which do not commute, the wedge product is not zero: $(\vec {\mathfrak A} \wedge \vec {\mathfrak A})_i = {\varepsilon_i}^{jk} \mathfrak A_j \mathfrak A_k = \frac{1}{2} {\varepsilon_i}^{jk}[\mathfrak A_j,\mathfrak A_k] \not=0$ ($\varepsilon$ being the Levi-Civita symbol).\\
We can show\cite{Viennot5} that:
\begin{itemize}
\item $\vec F = \tr(\rho_{a \alpha}^\epsilon \vec {\mathcal F})$ has the same physical meaning that in the previous sections (for closed quantum system). It is a magnetic field viewed in the world $\mathscr M_{a\alpha}$ by the particle. The crossings of the system eigenvalues $(\mu_b)$ appear as magnetic monopoles and ``wormholes'' between two worlds of the system. In addition to these usual structures of magnetic monopoles (points of singularity of $\vec F$), $\mathscr M_{a\alpha}$ presents also lines of strong magnetic field $\vec F$. These ``magnetic strings'' do not imply an exchange of worlds as the monopoles, but the passage of the path $\mathscr C$ through one of this line induces a sudden increase of the entropy (and a sudden increase of the entanglement between the system and the environment, see ref.\cite{Viennot5}).
\item $\vec B = \tr(\rho_{\alpha a}^\epsilon \vec {\mathcal B})$ is a measure of the entropy slow variation in the world $\mathscr M_{a\alpha}$ induced by variations of the classical parameters $\vec x$ (see ref.\cite{Viennot5}). Moreover, the crossings of the environment eigenvalues $(\nu_\beta)$ appear as singularities and ``wormholes'' between two worlds of the environment.
\end{itemize}
The presence of this $B$-field may seem strange in the magnetic analogy. In fact such a field appears in the context of string theory\cite{string} (we will discuss its meaning in string electrodynamics section VI). As the analogy with wormholes, this suggests a relation between the geometric phase theory and quantum gravity.

\section{Geometric phases in matrix models of quantum gravity}
Penrose argues that the many-worlds interpretation is unsatisfactory because it involves only standard quantum mechanics which does not take into account gravity\cite{Penrose}. This needs to consider a quantum gravity theory for which the lack of experimental data induces no evidence of its structure. The literature abounds of hypothetic candidates for the quantum gravity theory. The main approaches are \textit{string theory}\cite{string2}, \textit{quantum loop gravity}\cite{QLG}, \textit{non-commutative gravity}\cite{NCG} and \textit{emergent gravity}\cite{EmG}. This is not the place to make a review of this subject. We present here a simple model of quantum gravity in which the geometric phases have an interesting role.

\subsection{A matrix model of quantum gravity}
The main principle in the considered model is that quantum gravity must be obtained by considering the quantisation of the classical observables from the viewpoint of an ideal local Galilean observer (so a free falling observer). General relativity theory can be inferred from the Einstein's elevator thought experiment: an observer watches an object in an elevator cabin whose cable is cut (we neglect all friction forces). Due to the equivalence principle (the inertial mass is equal to the gravitational mass), the acceleration of the three bodies (the observer, the object and the elevator cabin) does not depend on their masses : $m\vec a_{/\mathcal R} = m \vec E \Rightarrow \vec a_{/\mathcal R} = \vec E$ (where $\vec E$ is the gravitational field and $\mathcal R$ is any Galilean frame). So, in the elevator frame $\mathcal R_*$, the accelerations of the object and of the observer is zero: $\vec a_{/\mathcal R_*}= \vec 0$. The gravity is then locally erased in the free falling frame $\mathcal R_*$ (no experiment of the observer inside the elevator cabin can reveal the gravity), justifying to identify gravity to an inertial effect and the free falling frames to local Galilean frames (local because experiments on large scale - outside the elevator cabin - can reveal the gravity since $\vec E$ depends on the distance to the planet center). So, free falling worldlines correspond to dynamics without external forces. But these ones are not straight lines. To sustain the principle of least action, we claim that the free falling worldlines are geodesics of the spacetime (a geodesics being the line of minimal length joining two points). These geodesics being curves, the spacetime is curved.\\
We return now to the quantisation problem of the gravity. Because of the elevator experiment, an ideal local Galilean observer sees locally a flat spacetime (any curved manifold can be locally identified with its tangent space: in the neighbourhood of a point $x_0$ on $\mathscr M$, we can write for any observable $f(x)=f(x_0)+\frac{\partial f}{\partial x^i}(x^i-x^i_0)+\mathcal O(\|x-x_0\|^2)$ permitting to identify the points of the neighbourhood with the tangent vectors generated by $(\partial_i)$: $x\simeq x_0 + (x^i-x_0^i)\partial_i$). We quantise then the observables of this one. The flat spacetime viewed by the local Galilean observer can be endowed with a coordinate systems $(x^A)$ (with $x^0=ct$, $x^1=x$, $x^2=y$ and $x^3=z$) and with the Minkowski metric $c^2 ds^2=\eta_{AB} dx^A dx^B = c^2 dt^2 - dx^2-dy^2-dz^2$ (where $s$ is the proper time of an observed object moving in direction $(dx,dy,dz)$ during the time $dt$ measured by the Galilean observer's clock). $(x^A)$ are the fundamental observables of the spacetime, since the other observables can be written as functions of $(x^A)$. The spacetime quantisation consists then to a quantisation rule $x^i \leadsto X^i$ (defining a semi-classical Poisson structure, see ref.\cite{Schneiderbauer} for more details) where $X^i \in \mathcal L(\mathscr H)$ is an operator of some Hilbert space $\mathscr H$ of the spacetime. In a next section, with the example of quantum wormholes we present a procedure to obtain such operators by using an algebra of creation and annihilation operators. We choose to not quantise the time $t$ since this one corresponds to the clock of the local Galilean observer (and not a time of the observed object). Due to the quantisation, $[X^i,X^j] \not= 0$ and so we have Heisenberg uncertainty relations $\Delta X^i \Delta X^j \geq \frac{1}{2} |\langle [X^i,X^j]\rangle|$. The set of noncommutative coordinate observables $(X^i)_i$ defines a fuzzy space\cite{fuzzy}, a simple example of a noncommutative manifold (in that meaning, the model can be viewed as a model of noncommutative gravity). We can interpret the fuzzy space as follows. Consider the eigenbasis of $X^1$, by definition this observable can be rewritten in this basis as
\begin{equation}
  X^1 = \left(\begin{array}{cccc} x^1_1 & 0 & 0 & \\ 0 & x^1_2 & 0 & \\ 0 & 0 & x^1_3 & \\ & & & \ddots \end{array} \right)
\end{equation}
The interpretation is obvious, $(x^1_a)$ are the possible outputs of the measure of $X^1$. We can then think that the $a$-th eigenstate corresponds to a ``point'' of coordinate on the first axis equal to the value $x^1_a$. But because of $[X^1,X^2]\not=0$, $X^2$ is not diagonal in the eigenbasis of $X^1$:
\begin{equation}
  X^2 = \left(\begin{array}{cccc} x^2_{11} & x^2_{12} & x^2_{13} & \\ x^2_{21} & x^2_{22} & x^2_{23} & \\ x^2_{31} & x^2_{32} & x^2_{33} & \\ & & & \ddots \end{array} \right)
\end{equation}
We would like interpret $x^2_{aa}$ as the coordinate on the second axis of the ``point'' associated with $a$-th eigenstate of $X^1$. But what are the off-diagonal elements $x^2_{ab}$? In usual quantum mechanics, off-diagonal elements are couplings between states. $x^2_{ab}$ is then a coupling between the two ``points'' $a$ and $b$ in the second direction. What is a coupling between two points? From a classical viewpoint, we can interpret this one as a string linking the two points, the tension of this one transmitting energy from a point to another one. The model can be then also associated with string theory. In this one we say that $(X^i)$ describes a stack of $D0$-branes (the ``points'' corresponding to the string ends) linked by bosonic strings and forming a noncommutative $D2$-brane (the fuzzy space). The quantisation rule $x^i \leadsto X^i$ can be then borrowed to string theory, and more precisely to the BFSS (Banks-Fischler-Shenker-Susskind) matrix theory\cite{BFSS}. But note that the previous discussion is basis-dependent, in another basis the possible outputs of the coordinate measurements and the couplings change. The description by ``points'' is then not pertinent, and the quantum degrees of freedom (D0-branes and strings) are inseparable objects. The notion of points is replaced by the one of quantum state. In any state $|\Psi \rrangle \in \mathbb C^2 \otimes \mathscr H$, $\llangle \Psi|X^i|\Psi \rrangle$ is the mean value of the coordinate $i$ of the state $|\Psi \rrangle$ measured by the observer, with a dispersion $\Delta_\Psi X^i = \sqrt{\llangle \Psi|(X^i)^2|\Psi \rrangle - \llangle \Psi|X^i|\Psi \rrangle^2}$. $\mathbb C^2$ is the Hilbert space of a spin degree of freedom associated with the orientability of the fuzzy space. More precisely, $\llangle \Psi|\vec \sigma|\Psi \rrangle$ ($\vec \sigma = (\sigma^1,\sigma^2,\sigma^2)$ are the Pauli matrices) can be viewed as the mean value of a normal vector to the fuzzy space defining a local orientation (by the right hand rule) at the state $|\Psi \rrangle$.\\
The local Galilean observer observes a ``test particle'' (to reveal the gravity, as in Einstein general relativity theory, the observer must send a test particle and observes its geodesic). This one is supposed to be a massless $\frac{1}{2}$-spin fermionic particle (a spin is needed to observe precession effects (Thomas or de Sitter precessions); the particle is supposed to be massless in order to this one does not curve the spacetime at the Planck length). Since the test particle is in the hand of the observer, it is described by classical coordinates $(x^i)$ measured by this one. The quantum state $|\Psi \rrangle$, the quantum coordinate observables $(X^i)$ and the classical coordinates $(x^i$) of the test particle are related by the noncommutative Dirac equation providing by the BFSS theory\cite{BFSS}:
\begin{equation}\label{DiracEq}
  \ihbar |\dot \Psi \rrangle = \slashed D_{x(s)} |\Psi(s) \rrangle
\end{equation}
with
\begin{equation}
  \slashed D_x = \frac{m_Pc^2}{\ell_P} \sigma_i \otimes (X^i - x^i\, \id_{\mathscr H})
\end{equation}
$s$ being the proper time of the test particle, $m_P= \sqrt{\frac{\hbar c}{G}}$ and $\ell_P = \sqrt{\frac{\hbar G}{c^3}}$ being the Planck mass and the Planck length. In the string theory viewpoint, the test particle is in fact a fermionic string linking the fuzzy space (described by $(X^i)$) to a probe $D0$-brane (described by $(x^i)$). $X^i-x^i\, \id_{\mathscr H}$ is then the quantum observable of the distance between the two ends of the fermionic string in the direction $i$. $(\sigma^i)$ are the spin observables of the fermionic string, and so $\sigma_i \otimes (X^i-x^i\, \id_{\mathscr H}) = \vec \sigma \odot (\vec X-\vec x \, \id_{\mathscr H})$ is the inner product of the spin and of the string vector (the vector between the two string ends). Eq.(\ref{DiracEq}) is a Schr\"odinger-like equation. $\slashed D_x$ can be then viewed as a kind of Hamiltonian. More precisely, $\slashed D_x$ is the displacement energy observable, intuitively the tension energy of the probe fermionic string. Indeed we have
\begin{equation}
  \slashed D_x^2 \propto \|\vec X - \vec x \, \id_{\mathscr H}\|^2 + \frac{\imath}{2} {\varepsilon_{ij}}^k \sigma_k \otimes [X^i,X^j]
\end{equation}
If the probe D0-brane (the test particle) is move far away from the fuzzy space, the distance $\|\vec X - \vec x\, \id_{\mathscr H}\|^2$ grows and the tension energy of the string increases. In a state $|\Psi\rrangle$ corresponding to a high delocalisation of the string end attached to the fuzzy space, we have $\Delta_\Psi X^i \Delta_\Psi X^j \geq \frac{1}{2} |\llangle \Psi|[X^i,X^j]|\Psi \rrangle|$ large. Due to the high delocalisation of the string end, this one has a large tension energy.\\
Where is gravity in this model? We can prove\cite{Klammer, Steinacker, Kunter} that at the thermodynamical limit (number of strings tending to infinity) and at the semi-classical limit ($\hbar \to 0$, limit of the macroscopic scale), gravity (curvature of spacetime) emerges at the macroscopic scale from the noncommutativity at the Planck scale of the quantum flat spacetime described by the fuzzy space. We can intuitively understand how curvature emerges from the noncommutativity. The main manifestation of a classical curvature can be found in the holonomy of the parallel transport of a tangent vector (a tangent vector parallel transported along a closed path does not return to itself after the transport). At the infinitesimal representation, this comes from the non-triviality of the operator $\nabla_i \nabla_j - \nabla_j \nabla_i$ where $\nabla_i$ stands for the covariant derivative in the direction $x^i$ (a vector field $\vec w$ is parallel transported along $\vec v$ if $v^i \nabla_i w^j = 0$, where $\nabla_i w^j = \frac{\partial}{\partial x^i} w^j + \Gamma^j_{ik} w^k$, $\Gamma^j_{ik}$ being the Christoffel symbols of the curved manifold). More precisely, for any co-tangent vector $v$, $(\nabla_i \nabla_j - \nabla_j \nabla_i) v_k = -\frac{1}{2} {R_{ijk}}^l v_l$ where ${R_{ijk}}^l$ is the Riemann curvature tensor. On a flat classical manifold, where the covariant derivatives are simply $\nabla_i = \frac{\partial}{\partial x^i}$, we have $\nabla_i \nabla_j - \nabla_j \nabla_i = 0$. In the case of a noncommutative space, where $x^i$ is replaced by an operator $X^i$, the natural derivative replacing $\frac{\partial}{\partial x^i}$ is the partial commutator $L_{X^i} = -\frac{\imath}{\ell_P^2} [X^i, \bullet]$. But in that case $L_{X^i} L_{X^j} - L_{X^j} L_{X^i} = L_{-\frac{\imath}{\ell_P^2} [X^i,X^j]}$, and so since $[X^i,X^j] \not=0$, the operator $L_{X^i} L_{X^j} - L_{X^j} L_{X^i} \not=0$ is not trivial (even if no Christoffel symbols are added to the natural derivatives). We can then understand the way how gravity/curvature emerges from the noncommutativity at the Planck scale of the flat quantum spacetime. In this meaning, the present model is also an example of emergent gravity theory. In this paper, we want to consider the emergence of gravity at the Planck scale, where the geometric phase associated with eq.(\ref{DiracEq}) has an important role.

\subsection{Emergent gravity by adiabatic quasi-coherent picture}
We want to consider the adiabatic regime of eq.(\ref{DiracEq}). We need then to examine the instantaneous eigenvectors of $\slashed D_x$:
\begin{equation}
  \slashed D_x |a,x\rrangle = \lambda_a(x)|a,x\rrangle
\end{equation}
particularly, we are interested by the state $|0,x\rrangle$ and a surface $\mathscr M \subset \mathbb R^3$ such that
\begin{equation}
 \forall x\in \mathscr M, \quad |\lambda_0(x)|<|\lambda_b(x)| \quad (\forall b\not=0)
\end{equation}
$|0,x\rrangle$ minimises then the displacement energy (the state for which the probe fermionic string has zero tension energy $\lambda_0(x)=0$ is such that the two ends are more close as possible, and the probe D0 brane - the test particle - is more close as possible to the fuzzy space). Moreover, if $\lambda_0(x)=0$ we can prove\cite{Schneiderbauer} that $|0,x\rrangle$ minimises the Heisenberg uncertainty relations and is then an analogue in quantum gravity matrix model to a coherent state\cite{Perelomov}. So, $|0,x\rrangle$ are the states closest to classical states of a classical manifold, and then closest to the usual notion of points of a manifold. But they are not true classical points (Dirac distributions) since the non-separability of the things that we would like to call ``points'' of the fuzzy space is sustained by the non-orthogonality of the states $\llangle 0,y|0,x\rrangle \not=0$ even if $x\not=y$. So being at a ``quantum point'' $|0,x\rrangle$ induces to have a non-zero probability of measuring the ``location'' at $|0,y\rrangle$ for any $y$. $|0,x\rrangle$ is a state of the fuzzy space which is labelled by a point $x$ in the classical space being in the observer's mind ($x$ being the location of the test particle in the flat classical space that the observer thinks he sees). The ``quantum reality'' from the viewpoint of the test quantum particle in a quantum space is a state $|0,x\rrangle$ non-separated from the others (it is only separated from - orthogonal to - the states of higher displacement energies). $|0,x\rrangle$ is called quasi-coherent state when $\mathscr M = \{x \in \mathbb R^3, \det(\slashed D_x) = 0\}$. $\mathscr M$ is then the classical surface closest to the ``quantum geometry'' defined by the fuzzy space associated with $(X^i)$. $\mathscr M$ can be viewed as a slice of space representing the emergent geometry at the Planck scale in the quasi-coherent state. In general, $\mathscr M$ is a curved surface inducing an emergent curvature and then inducing gravity effects. $\llangle 0,x|\vec \sigma|0,x\rrangle$ (with $x \in \mathscr M$) is a normal vector at $\mathscr M$ at the point $x$. $\mathscr M$ is endowed with the metric induced by the embedding of $\mathscr M$ in $\mathbb R^3$:
\begin{equation}
  \gamma_{ab} = \delta_{ij} \frac{\partial x^i}{\partial u^a} \frac{\partial x^j}{\partial u^b}
\end{equation}
where $(u^1,u^2)$ are local curvilinear coordinates on $\mathscr M$. We can prove\cite{Viennot6} that $\gamma_{ab} = \gamma_{ab}^{dist} + \gamma_{ab}^{nc}$ with $\gamma^{dist} = (\partial_a \llangle 0,x|) \|\vec X-\vec x\, \id_{\mathscr H}\|^2 \partial_b|0,x\rrangle du^a du^b$ is the quadratic variation of the mean value of the square distance observable. The main contribution to the metric is due to the noncommutativity of the fuzzy space $\gamma^{nc} = \frac{1}{4} (\partial_a\llangle 0,x|) [\sigma_i,\sigma_j] \otimes [X^i,X^j] \partial_b|0,x\rrangle du^adu^b$. The curvature (the non-trivial form of $\gamma$) emerges then well from the noncommutativity of the quantum space.\\
In the usual adiabatic approximation, the solution of eq.(\ref{DiracEq}) with $|\Psi(0) \rrangle = |0,x(0)\rrangle$ is
\begin{equation}
  |\Psi(s)\rrangle \simeq e^{-\imath \int_0^s A_i(x(s')) \dot x^i(s') ds'} |0,x(s)\rrangle
\end{equation}
for a path $s \mapsto x(s) \in \mathscr M$ on the slice of space $\mathscr M$ and with
\begin{eqnarray}
  \vec A(x) & = & -\imath \llangle 0,x|\vec \nabla|0,x\rrangle_{|\mathscr M} \\
  & = & -\imath \llangle 0,x|\frac{\partial}{\partial x^i}|0,x\rrangle_{|x=x(u)} \frac{\partial x^i}{\partial u^a} \vec e_a
\end{eqnarray}
$\vec e_a$ being the unit vector tangent to $\mathscr M$ in the direction $u^a$ at $x(u)$. What is the role of $\vec A$ in quantum gravity? Firstly let $\mathscr M^{(s)}$ be the copy of $\mathscr M$ viewed by the test particle at the proper time $s$. $\mathcal M = \bigsqcup_s (s,\mathscr M^{(s)}) \subset \mathbb R^{4}$ is then a three-dimensional submanifold of the classical spacetime of the observer, foliated with respect to the proper time (the leaf $\mathscr M^{(s)}$ is a two dimensional slice of space at fixed proper time). Suppose that $\mathscr M^{(s)}$ is associated with $|0,x\rrangle$, then $\mathscr M^{(s+ds)}$ is associated with $e^{- \imath \vec A \cdot \frac{d\vec x}{ds} ds} |0,x\rrangle$. Since $|0,x\rrangle$ is the quantum state of the fuzzy space closest to a point of a classical manifold, we see that starting from a state $|0,x\rrangle$ at time $s$, we arrive at the state $e^{- \imath \vec A \cdot \frac{d\vec x}{ds} ds} |0,x\rrangle$ at time $s+ds$ and not at the equivalent state $|0,x\rrangle$. $\vec A$ defines the shift (the geometric phase) between the two states. Then, in the classical point of view, $\vec A$ can be identified with the shift vector of the spacetime foliation (let $P$ be the point on the leaf $\mathscr M^{(s)}$ of curvilinear coordinates $u$, $P'$ be the point at the intersection of the normal vector at $\mathscr M^{(s)}$ at $P$ and $\mathscr M^{(s+ds)}$ and $P''$ be the point of $\mathscr M^{(s+ds)}$ of curvilinear coordinates $u$, then the shift vector is $\vec A ds = \overrightarrow{P'P''}$). We can prove\cite{Viennot6} that the metric of the emergent spacetime $\mathcal M$ is
\begin{equation}\label{emergentmetric}
  c^2ds^2 = c^2dt^2 -(A^a cdt + du^a)(A^b cdt+du^b)\gamma_{ab}
\end{equation}
which is well the metric of a foliated spacetime of shift vector $\vec A$. We can see that the preferred basis of the adiabatic picture induces a preferred foliation of the spacetime $\mathcal M$ (by defining the metric and the shift vector via the state $|0,x\rrangle$). This preferred foliation appears at the search of the classical spacetime closest to the quantum one, and so the search of the most intelligible representation for the observer at the vicinity of this observer (we join here the argument already evoked at the end of section III). The link between preferred bases of the Everett interpretation and preferred foliations in relativity has been already point out in ref.\cite{Wallace2} by philosophical arguments. The usual interpretation of $\vec A$ is a magnetic potential in $\mathscr M$. What is the meaning of the fact that this magnetic potential is identified with the shift vector? As shown in ref.\cite{Viennot6}, the presence of $\vec A$ in the emergent metric eq.(\ref{emergentmetric}) induces that the geodesics in $\mathcal M$ are curved in the same way as trajectories of a charged particle moving in a magnetic field. So the usual interpretation of $\vec A$ and its role in emergent gravity are consistent. More precisely, it is shown in ref.\cite{Viennot6} that $\vec A$ is a manifestation of the torsion of $\mathcal M$. In contrast with usual general relativity, the emergent spacetime at Planck scale is not torsion free (an effect also present at the semi-classical thermodynamical limit\cite{Steinacker2}). This torsion is in the emergent classical geometry the souvenir of the quantum aspect of the spacetime at the Planck scale.\\
We can also consider the weak adiabatic transport, in that case $\vec {\mathfrak A}$ is associated with the entanglement between the spin state of the test particle and the quantum state of the fuzzy space. $-\imath \vec {\mathfrak A}^{off}$ (with $\vec {\mathfrak A}^{off} = \vec {\mathfrak A} - \frac{1}{2} \tr \vec {\mathfrak A}$) is the Lorentz connection of the emergent spacetime $\mathcal M$, so the observable governing the spin precession\cite{Viennot6}.\\

Is this approach of quantum gravity consistent with the Everett's many-worlds interpretation? To answer to this question, it is interesting to consider the example of a quantum wormhole.

\subsection{Quantum wormholes}
\begin{figure}
  \includegraphics[width=7cm]{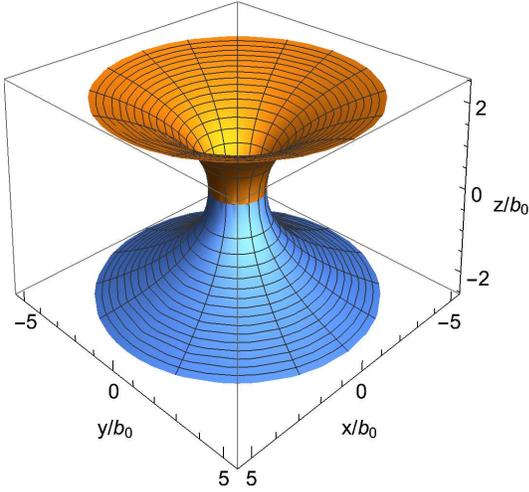}
  \caption{\label{MTwormhole} Space slice at $\theta=\frac{\pi}{2}$ and $t=C^{st}$ of a Morris-Thorne wormhole of throat radius $b_0$ embedded in $\mathbb R^3$. The sheets of space $\pm z(r)$ linked by the wormhole are represented by two different colours. The wormhole itself is the circle of radius $b_0$ at which the two sheets are glued together (by varying $\theta$, this one becomes a sphere of radius $b_0$).}
\end{figure}

The concept of wormhole in general relativity initially comes from the Einstein-Rosen bridges\cite{Einstein, Flamm} which are extensions of the black hole solutions of the Einstein equations. This kind of wormhole is not traversable. The quantum version of these ones by a fuzzy space has been studied in ref.\cite{Viennot7} where it is shown that a particle can traverse a quantum Einstein-Rosen bridge by tunnelling effect. Here we want to consider another kind of wormholes, the Morris-Thorne wormholes\cite{Morris} which are classically traversable. A classical Morris-Thorne wormhole is characterised by the following spacetime metric:
\begin{equation}\label{WHmetric}
  c^2ds^2 = c^2dt^2 - d\ell^2-(b_0^2+\ell^2)(d\theta^2+\sin^2\theta d\varphi^2)
\end{equation}
where $b_0$ is the radius of the wormhole throat, $\ell$ is the proper radial distance to this throat. The space slice of equation $\theta=\frac{\pi}{2}$ and $t=C^{st}$ can be embedded in $\mathbb R^3$ by
\begin{equation}
  z(r) = b_0 \ln \left(\frac{r}{b_0} +\sqrt{\frac{r^2}{b_0^2}-1} \right)
\end{equation}
with $r=\sqrt{\ell^2+b_0^2}$, $(r,\varphi)$ being polar coordinates of the $(x,y)$-plane. The slice is represented fig.\ref{MTwormhole}.
To define a quantum Morris-Thorne wormhole, we follow the procedure used in ref.\cite{Viennot7}, by considering the quantum wormhole as a deformation of the noncommutative plane. Let $a$ and $a^+$ be the annihilation and creation operators onto the Fock space $\mathscr H$ (the operators of the quantum harmonic oscillator). The quantum coordinate observables $X$ and $Y$ are defined by $X=\ell_P \frac{(a+a^+)}{2}$ and $Y=\ell_P \frac{(a-a^+)}{2\imath}$. In other words, $a = \frac{1}{\ell_P} (X+\imath Y)$ is a quantum complex coordinate observable onto the plane. The quantum radial coordinate observable is defined by
\begin{equation}
  \hat r = \ell_P \sqrt{a^+a} = \ell_P \sum_{n=0}^{+\infty} \sqrt{n} |n\rangle \langle n|
\end{equation}
where $(|n\rangle)$ is the canonical basis of the Fock space $\mathscr H$ ($a^+|n\rangle = \sqrt{n+1}|n+1\rangle$ and $a|n\rangle = n|n-1\rangle$ except for $n=0$ where $a|0\rangle = 0$). The last quantum coordinate observable is then defined by
\begin{eqnarray}
  Z & = & z(\hat r) \\
  & = &  \ell_P \sum_{n=0}^{+\infty} \ln\left(\sqrt n + \sqrt{n-1} \right) |n\rangle \langle n|
\end{eqnarray}
where we have chosen to consider a quantum wormhole with throat radius equal to the Planck length $b_0=\ell_P$. For one sheet, the Dirac operator (the displacement energy observable) is
\begin{eqnarray}
  \slashed D_x & = & \frac{m_Pc^2}{\ell_P} \sigma_i \otimes (X^i - x^i) \\
  & = & m_Pc^2 \left(\begin{array}{cc} (Z - z)/\ell_P & a^+ - \bar \alpha \\ a - \alpha & (- Z^\dagger + \bar z)/\ell_P \end{array} \right)
\end{eqnarray}
with $\alpha = (x+\imath y)/\ell_P$ and $z=\ell_P \ln(|\alpha|+\sqrt{|\alpha|^2-1})$ and where the matrix is written in the canonical basis of $\mathbb C^2$ for which $\sigma_3|\frac{1}{2}\rangle=|\frac{1}{2}\rangle$ and $\sigma_3|-\frac{1}{2}\rangle = - |-\frac{1}{2}\rangle$.\\

\begin{figure}
  \includegraphics[width=7cm]{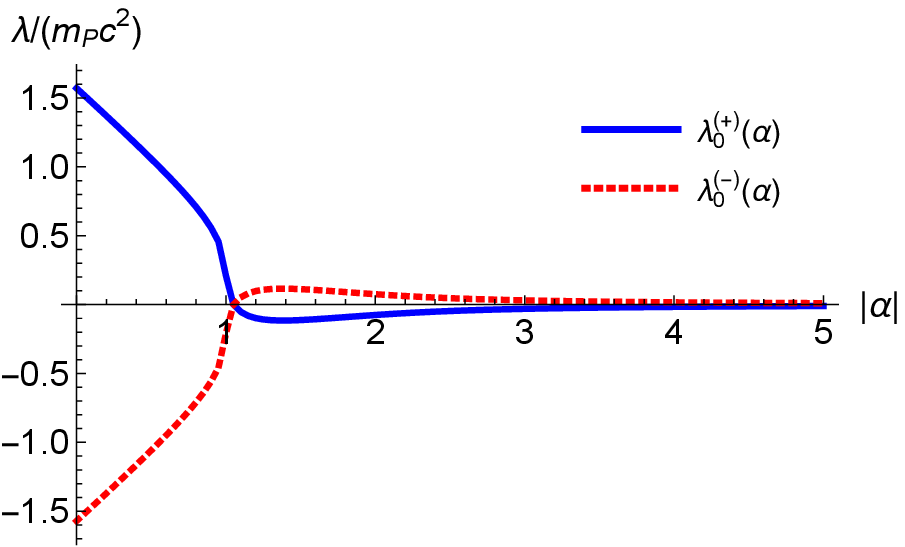}\\
  \includegraphics[width=7cm]{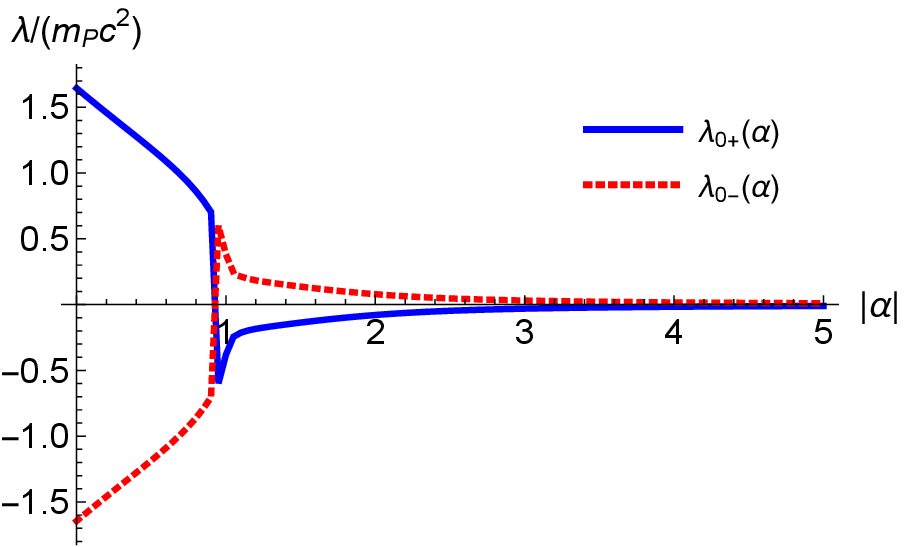}\\
  \includegraphics[width=7cm]{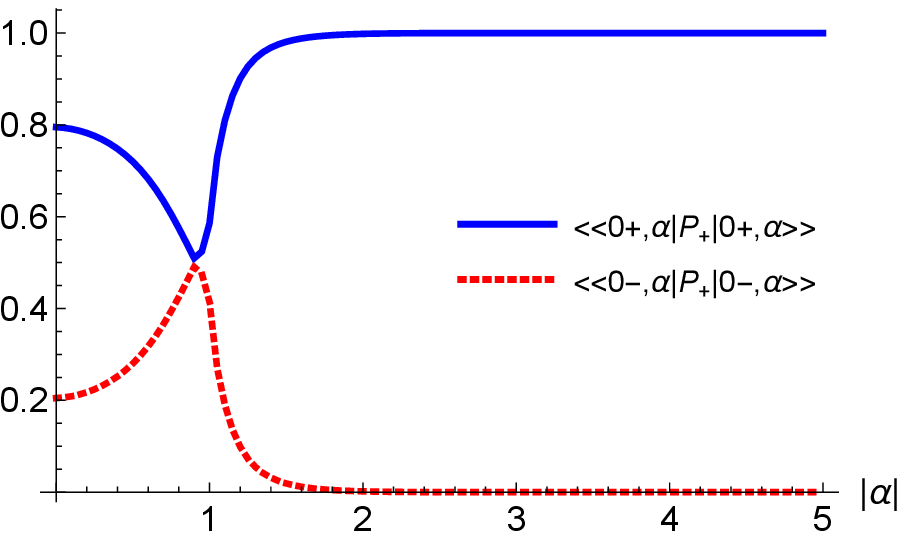}\\
  \includegraphics[width=7cm]{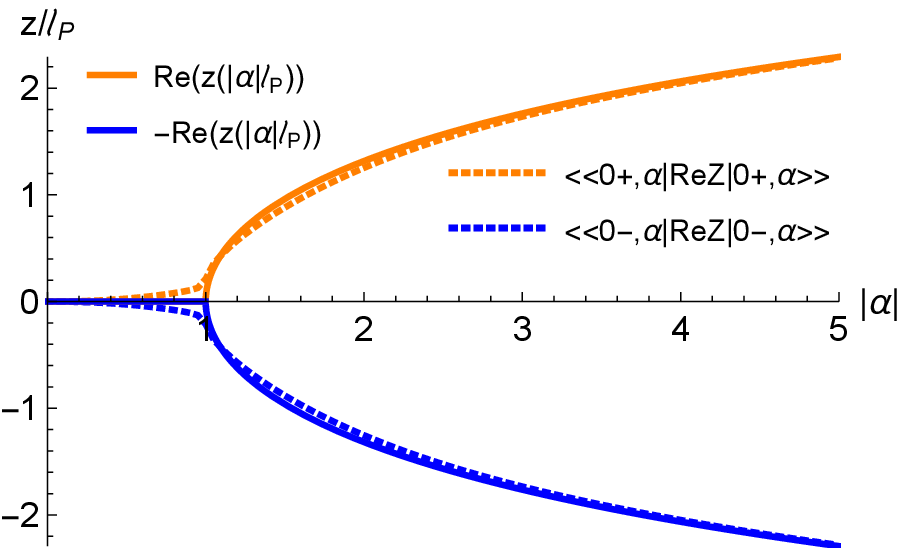}\\
  \caption{\label{QEWH} Up: minimal displacement energies of the single sheet operators: $|\lambda^{(\pm)}_0(x)| = \min |\mathrm{Sp}(\slashed D_x^{\pm})|$. Middle (first): minimal displacement energies of the double sheet operator, solutions of eq.(\ref{EigenEqWH}). Middle (second): Probability that the test particle is on the upper sheet when its state is $|0\pm,\alpha\rrangle$ (solutions of eq.(\ref{EigenEqWH})). Down: Comparison of the profil of the classical wormhole slice fig.\ref{MTwormhole} with the mean values of $\Re\mathrm{e}(Z)$ in the eigenstates $|0\pm,\alpha\rrangle$.}
\end{figure}

%With $Z=0$, $\slashed D^{Z=0}_x$ is the Dirac operator of a noncommutative plane. Their eigenvectors are
%\begin{eqnarray}
%  \slashed D_x^{Z=0} |0,\alpha \rrangle^{Z=0} & = & 0 \\
%  \slashed D_x^{Z=0} |(\pm,n),\alpha \rrangle^{Z=0} & = & \pm\sqrt n |(\pm,n),\alpha\rrangle^{Z=0}
%\end{eqnarray}
%with $|0,\alpha\rrangle^{Z=0} = |0\rangle \otimes |\alpha\rangle$, $|0\rangle$ being the eigenstate of $\sigma_3$ and $|\alpha\rangle$ being the usual coherent state of an harmonic oscillator\cite{Perelomov}
The quasi-coherent states are similar to the usual Perelomov coherent states\cite{Perelomov}, these ones are: 
\begin{equation}
  |\alpha \rangle = e^{-|\alpha|^2/2} \sum_{n=0}^{+\infty} \frac{\alpha^n}{\sqrt{n!}} |n\rangle
\end{equation}
The Heisenberg uncertainty relation for $X$ and $Y$ is minimised in the state $|\alpha\rangle$.
%The other eigenstates are $|(\pm,n),\alpha\rangle^{Z=0} = \frac{1}{\sqrt 2}(|0\rangle \otimes |n\rangle_\alpha \pm |1\rangle \otimes |n-1\rangle_\alpha)$ with
%\begin{equation}
%  |n\rangle_\alpha = \frac{(a^+-\bar \alpha)^n}{\sqrt{n!}}|\alpha\rangle
%\end{equation}
%$|0,\alpha\rrangle^{Z=0}$ is the quasi-coherent state of the noncommutative plane, it defines the simple manifold $M^{Z=0} = \{(\ell_P \Re\mathrm{e}(\alpha), \ell_P \Im\mathrm{m}(\alpha),0), \alpha \in \mathbb C\} \subset \mathbb R^3$, the plane $z=0$. The excited states $|(\pm,n),\alpha\rrangle^{Z=0}$ can be viewed as states of local deformation at the point of complex coordinate $\alpha$ (positive or negative deformation of magnitude $\sqrt n$)\cite{Viennot7}. It is natural to represent the quasi-coherent state of $\slashed D_x^{(\pm)}$ onto this basis.\\
We note that the points $|\alpha|<1$ ($\iff r < b_0=\ell_P$) are forbidden in the classical wormhole (the ``hole'' - the genus - in fig.\ref{MTwormhole}). These points are forbidden since they do not belong to the spacetime (a coordinate $r<b_0$ is not in the slice of space). But in the quantum model, this implies that $Z$ is not self-adjoint:
\begin{eqnarray}
  Z/\ell_P & = & \sum_{n=1}^{+\infty} \ln\left(\sqrt n + \sqrt{n-1} \right) |n\rangle \langle n| \nonumber \\
  & & - \imath \frac{\pi}{2} |0\rangle \langle 0|
\end{eqnarray}
where we have chosen the Riemann sheet of the complex square root $\sqrt{-1} = - \imath$ in order to have a negative imaginary part (moreover we have chosen the main value of the complex logarithm). Even if the region $|\alpha|<1$ is classically forbidden, quantum wave functions can penetrate in this region as evanescent waves.
\begin{equation}
  \langle \alpha|\Im\mathrm{m} Z|\alpha\rangle = - \frac{\pi}{2} \ell_P  e^{-|\alpha|^2}
\end{equation}
For large value $|\alpha|\gg 1$, this quantity is negligible and no effect of the nonselfadjointness are visible. For small value $|\alpha|\ll 1$, we have $\langle \alpha|\Im\mathrm{m} Z|\alpha\rangle \simeq -\frac{\pi}{2} \ell_P$, and it follows
\begin{equation}
  e^{-\ihbar^{-1} \frac{m_P c^2}{\ell_P} \imath \Im\mathrm{m} Z s} |\alpha\rangle \simeq e^{- \frac{m_pc^2}{\hbar} \frac{\pi}{2} s} |\alpha \rangle
\end{equation}
During the evolution (generated by the eq.(\ref{DiracEq})) a state $|\alpha\rangle$ (with $|\alpha|\ll 1$) in the forbidden region is exponentially dissipated (as expected for a evanescent state). The characteristic time of the dissipation is $\frac{2}{\pi} t_P$ where $t_P = \frac{\hbar}{m_P c^2} = \sqrt{\frac{\hbar G}{c^5}}$ is the Planck time. So wave function entering in the classically forbidden region is quickly dissipated. But this dissipation is only present in this single sheet model. Where does the wave function disappear? In the other sheet of the representation of fig.\ref{MTwormhole} (by definition the wormhole is a transition between two spacetime sheets). A double sheet quantum model is represented by the following Dirac operator:
\begin{equation}
  \slashed{D}_x^{\rotatebox{90}{$\scriptstyle \bowtie$}} = m_Pc^2 \left(\begin{array}{cc} \slashed{D}_x^+/(m_Pc^2) & \imath \Im\mathrm{m}(Z-z)/\ell_P \\ -\imath \Im\mathrm{m}(Z-z)/\ell_P & \slashed{D}_x^-/(m_Pc^2) \end{array} \right)
\end{equation}
with
\begin{equation}
  \slashed D_x^\pm = m_Pc^2 \left(\begin{array}{cc} \pm \Re\mathrm{e}(Z-z)/\ell_P & a^+ - \bar \alpha \\ a - \alpha & \mp \Re\mathrm{e}(Z-z)/\ell_P \end{array} \right)
\end{equation}
$\Re\mathrm{e}(Z)=\ell_P \sum_{n=1}^{+\infty} \ln\left(\sqrt n + \sqrt{n-1} \right) |n\rangle \langle n|$. The matrix $\slashed{D}_x^{\rotatebox{90}{$\scriptstyle \bowtie$}}$ is written in the basis $(|+\rangle,|-\rangle)$ of the states of presence in the two sheets. The dissipation operator of the single sheet representation, $\imath \Im\mathrm{m}(Z)=-\imath \frac{\pi}{2}|0\rangle \langle 0|$, is now treated as a coupling between the states $|\pm\rangle$ (when the wave function disappears from the upper sheet, it goes to the lower one). We solve the eigenequation:
\begin{equation}\label{EigenEqWH}
  \slashed{D}_x^{\rotatebox{90}{$\scriptstyle \bowtie$}}|0\pm,\alpha\rrangle = \lambda_{0\pm}(\alpha)|0\pm,\alpha\rrangle
\end{equation}
with $|\lambda_{0\pm}(\alpha)| = \min |\mathrm{Sp}(\slashed D_x)|$. We have two minimal displacement energies ($\lambda_{0-} = - \lambda_{0+}$) due to the degenerescence induced by the double sheet. The sheet label $\pm$ is attributed with respect to the probabilities to be on a sheet, more precisely the label $+$ is attributed such that $\langle 0+,\alpha|P_+|0+,\alpha\rangle > \langle 0+,\alpha|P_-|0+,\alpha\rangle$ where $P_\pm = |\pm\rangle \langle \pm|$. The result of the numerical solving of eq.(\ref{EigenEqWH}) is shown fig.\ref{QEWH}.
As expected, an eigenvalue crossing occurs in the neighbourhood of the classical throat of the wormhole. This crossing induces a transition from an eigenstate associated with the upper sheet to an eigenstate associated with the lower sheet. But it is interesting to note that at the throat the eigenstates correspond to an equiprobability to be on the two sheets. Classically, the border between the two sheets is arbitrary chosen at the minimal throat, but in fact the two sheets form a connected surface (fig.\ref{MTwormhole}), and other choices are possible. Quantically, this is replaced by a quantum superposition of the states $|\pm\rangle$ in the neighbourhood of the throat for the eigenstate $|0+,\alpha\rangle$ (more precisely the state $|0+,\alpha\rangle$ presents entanglements between the states $|\pm\rangle$ and the states of the other quantum degrees of freedom). Finally we can claim that a quantum wormhole in the meaning of emergent gravity is well a ``many-world wormhole'' of the interpretation of the quantum adiabatic dynamics.\\
Remark: fig.\ref{QEWH} shows also that the minimal displacement energies asymptotically cross at $|\alpha|\to +\infty$, but this crossing cannot induce transition since the coupling is zero $\lim_{|\alpha|\to+\infty} \llangle 0+,\alpha|\Im\mathrm{m}(Z)|0-,\alpha \rrangle = 0$.\\

The computation of $\vec A^\pm = -\imath \ell_P^{-1} \llangle 0\pm,\alpha|\vec \nabla_\alpha|0\pm,\alpha \rrangle$ is not obvious, but by using a decomposition onto the eigenstates of the noncommutative plane\cite{Viennot6, Viennot7}, we can write that $\vec A^\pm = \frac{r}{\ell_P} \vec e_\varphi + A_r^\pm(r) \vec e_r$ with $A_r^\pm$ going quickly to $0$ with $r\gg 1$. The metric of the emergent spacetime is then
%\begin{eqnarray}
%  c^2ds^2 & = & \left(1-(\ell^2/\ell_P^2+1)+\frac{|A_r^+|^2}{1+\ell_P^2/\ell^2}\right) c^2dt^2 \nonumber\\
%  & & - 2 (\ell^2/\ell_P^2+1)\ell_P d\varphi cdt - \frac{2 A_r^+}{\sqrt{1+\ell_P^2/\ell^2}} d\ell cdt \nonumber\\
%  & & - d\ell^2 - (\ell^2+\ell_P^2) d\varphi^2
%\end{eqnarray}
\begin{eqnarray}
  c^2ds^2 & = & c^2dt^2 \nonumber\\
  & & - (\ell^2/\ell_P^2+1)(cdt+\ell_P d\varphi)^2 \nonumber\\
  & & - \left(\frac{A_r^\pm}{\sqrt{1+\ell_P^2/\ell^2}}cdt+d\ell\right)^2
\end{eqnarray}
with $\ell=\sqrt{r^2-\ell_P^2}$. The difference with respect to the classical metric eq.(\ref{WHmetric}) provides from the shift vector $\vec A^\pm$, the manifestation of the torsion at the Planck scale.

\section{Emergence and complexity}
Emergentism is a philosophical idea which opposes to reductionism by thinking that fundamental new laws emerge from the complexity\cite{Anderson}. In other words, the behaviour of a complex system cannot completely be reduced to the individual behaviours of its parts. This concept is particularly important in statistical physics where the complexity results from the large number of degrees of freedom. But the complexity can also result from nonlinearities, and unexpected organised behaviours can emerge from simple laws which at first appear to give only disorderly behaviours, as for example the Langton's ant cellular automaton\cite{Langton}. We have previously noted that the many-worlds are an emergent structure (and not a predefined spacetime property) coming from the diagonalisation of $H$ (the potential measurement of energy by the observer). One could say that for an unobserved system, the many-worlds do not exist and emerge only at the observation. The complexity coming from the bipartite interaction between the system and the observer. This is in agreement with the idea that the preferred basis of the many-worlds picture comes from decoherence processes resulting from entanglement of the system with a large environment. The complexity is then the same than in statistical physics (the large number of degrees of freedom of the environment). In the previous section, we have presented a simple model of quantum spacetime resulting from the quantisation of the classical flat spacetime, in which gravity emerges at the macroscopic scale with the thermodynamical limit (anew an emergence as in statistical physics). We have presented also the emergence of gravity at the Planck scale in the adiabatic quasi-coherent picture which results from the observation of a test particle whose the spin state is entangled with the spacetime state (the state $|0,x\rrangle$ is generally entangled). An emergence in the same way than the many-worlds structure emergence.\\

In the adiabatic picture, the emergence of the many worlds $\{\mathscr M_a\}_a$ is accompanied by the emergence of gauge fields $\{\vec A^a\}$ generators of the geometric phases. Does this emergence reflect the complexity of the problem? To answer to this question, it needs to discuss the complexity of a gauge theory. It is interesting to consider first the simplest example of electrostatics. Let a first observer and $U^\alpha$ be its neighbourhood in the space. This observer defines (by observations/measurements) an electric potential $V^\alpha$ and an electric field $\vec E = -\grad V^\alpha$. Let a second observer and its neighbourhood $U^\beta$ who defines the fields $V^\beta$ and $\vec E = -\grad V^\beta$. Suppose that the two neighbourhoods have a non zero intersection. On $U^\alpha \cap U^\beta$ the two observers are agree concerning the electric field, but due to the arbitrary gauge choice, in general we have $V^\beta(x)-V^\alpha(x) = k^{\alpha \beta} \in \mathbb R$ ($\forall x \in U^\alpha \cap U^\beta$). To take into account this question of the gauge choice, we can claim that the electrostatic gauge theory is defined by the following equivalence class of potentials $[V]_{\mathcal D} = \{V^\alpha + k^\alpha, \text{ with }  \forall x \in U^\alpha \cap U^\beta, V^\beta(x) - V^\alpha(x) = C^{ste}\}_{k^\alpha \in \mathbb R, \{U^\alpha\}_\alpha}$, the choice of the good open cover\footnote{A good open cover $\{U^\alpha\}$ of a manifold $\mathscr M$, is a set of open simply connected subsets of $\mathscr M$ (one piece no hole), such that $\bigcup_\alpha U^\alpha = \mathscr M$ and such that any intersection of a finite number of open sets $\{U^\alpha\}$ is also a simply connected open set.} of the space $\{U^\alpha\}_\alpha$ being arbitrary. The mathematicians called $[V]_{\mathcal D}$ a Deligne cohomology class\cite{Brylinski} of degree 1 (a concept of algebraic topology), it defines a gauge invariant field by $\vec E = - \grad V$.\\
Consider now the case of magnetism. Now the two observers are not agree concerning the magnetic potential: $\vec A^\beta - \vec A^\alpha = \grad \chi^{\alpha \beta}$ where $\chi^{\alpha \beta}$ is a function on $U^\alpha \cap U^\beta$. Generally, $\forall x \in U^\alpha \cap U^\beta \cap U^\gamma$, we have $\chi^{\alpha \gamma} = \chi^{\alpha \beta} + \chi^{\beta \gamma}$. But this is not the case in presence of a magnetic monopole where $\chi^{\alpha \beta} + \chi^{\beta \gamma} - \chi^{\alpha \gamma} = z^{\alpha \beta \gamma} \in \mathbb Z$ for which $\{U^\alpha,U^\beta,U^\gamma\}$ is a good open cover of a sphere centred onto the monopole and $z^{\alpha \beta \gamma}$ is its quantised magnetic charge (see ref.\cite{Frankel}). Finally the magnetic gauge theory is defined by the degree 2 Deligne cohomology class $[\chi,\vec A]_{\mathcal D} = \{(\chi^{\alpha \beta} + \zeta^\beta-\zeta^\alpha,\vec A^\alpha + \grad \zeta^\alpha), \text{ with } \forall x \in U^\alpha \cap U^\beta, \vec A^\beta - \vec A^\alpha = \grad \chi^{\alpha \beta} \text{ and } \forall x \in U^\alpha \cap U^\beta \cap U^\gamma, \chi^{\beta \gamma} - \chi^{\alpha \gamma} + \chi^{\alpha \gamma} = C^{ste} \}_{\zeta^\alpha \in \underline{\mathbb R}_{U^\alpha}, \{U^\alpha\}_\alpha}$  defining a gauge invariant field $\vec F = \rot \vec A$ ($\underline X_U$ denoting the set of differentiable functions from $U$ to $X$). The electromagnetic theory in the Minkowski spacetime is also a degree 2 Deligne class with the potential four-vector in place of $\vec A$ and the Faraday tensor in place of $\vec F$. The interest to view the Deligne class $[\chi,\vec A]_{\mathcal D}$ as the fundamental essence of the magnetism is related to the Aharonov-Bohm effect\cite{Aharonov}. In this effect, a particle which sees only a zero magnetic field in the whole of its path, presents a measurable effect (by interferences) of the non-zero magnetic potential. We could then think that this one is more pertinent to be the essence of magnetism in place of the magnetic field. But $\vec A$ is submitted to the arbitrary gauge choice, whereas a physical essence of the Reality needs to be uniquely defined. The circulation of $\vec A$, $\oint_{\mathscr C} \vec A \cdot d\vec x$, has the default to be defined on non-local (extended) objects (closed paths $\mathscr C$) in place of the points of the space. The equivalence class $[\chi,\vec A]_{\mathcal D}$ seems then to be more consistent to be the essence of the theory\footnote{This is not unusual to consider equivalence classes as physical objects. This is for example the case of the ``wave functions'' in quantum mechanics. These ones are not simply square integrable functions, but an equivalence class of square integrable functions up to the addition of functions of zero Lebesgue measure (functions of zero integration), to have a true Hilbert space (the addition of zero Lebesgue measured function to a wave function does not change the probabilities nor the measurement outputs).}.\\
We can increment the construction and define a degree 3 Deligne cohomology class as being $[\xi,\vec \eta,\vec B]_{\mathcal D} = \{(\xi^{\alpha \beta \gamma} + \zeta^{\beta \gamma} - \zeta^{\alpha \gamma} + \zeta^{\alpha \beta}, \vec \eta^{\alpha \beta} - \grad \zeta^{\alpha \beta} + \vec k^\beta-\vec k^\alpha, \vec B^\alpha + \rot \vec k^\alpha) \text{ with } \forall x \in U^\alpha \cap U^\beta, \vec B^\beta - \vec B^\alpha = \rot \vec \eta^{\alpha \beta} \text{ and } \forall x \in U^\alpha \cap U^\beta \cap U^\gamma, \vec \eta^{\beta \gamma} - \vec \eta^{\alpha \gamma}+\vec \eta^{\alpha \beta} = - \grad \xi^{\alpha \beta \gamma} \text{ and } \forall x \in U^\alpha \cap U^\beta \cap U^\gamma \cap U^\delta, \xi^{\beta \gamma \delta} - \xi^{\alpha \gamma \delta} + \xi^{\alpha \beta \delta} - \xi^{\alpha \beta \gamma} = C^{ste} \}_{\zeta^{\alpha \beta} \in \underline{\mathbb R}_{U^\alpha \cap U^\beta}, \vec k^\alpha \in \underline{\mathbb R^3}_{U^\alpha}, \{U^\alpha\}_\alpha}$ defining a gauge invariant field $H = \Div \vec B^\alpha$. This class corresponds to a physical theory, the 2-form electrodynamics\cite{Henneaux}. In usual magnetism, a gauge invariant quantity is the potential vector circulation $\oint_{\mathscr C} \vec A \cdot d\vec x$ on a closed trajectory $\mathscr C$ (it is the fundamental quantity in the Aharonov-Bohm effect\cite{Aharonov}). For a path $\mathscr C$ crossing several open sets of different gauge choices, we have a gluing relation $\int_{\mathscr C} \vec A \cdot d\vec x = \int_{\mathscr C^\alpha} \vec A^\alpha \cdot d\vec x + \chi^{\alpha \beta}(x^{\alpha \beta}) + \int_{\mathscr C^\alpha} \vec A^\alpha \cdot d\vec x$, where $\mathscr C$ is cut into two parts, $\mathscr C^\alpha \subset U^\alpha$ and $\mathscr C^\beta \subset U^\beta$, at an arbitrary point $x^{\alpha \beta} \in \mathscr C \cap U^\alpha \cap U^\beta$ (the circulation is independent of this arbitrary choice). Suppose that we generalise the magnetism theory by replacing point particles by closed strings. In that case, the closed trajectory $\mathscr C$ is replaced by a closed surface $\mathscr T$ having the topology of a torus. The circulation must then be replaced by a flux $\oiint_{\mathscr T} \vec B \cdot d\vec S$. For a surface $\mathscr T$ having the topology of a tube crossing two open sets, we have now the gluing relation $\iint_{\mathscr T} \vec B \cdot d\vec S = \iint_{\mathscr T^\alpha} \vec B^\alpha \cdot d\vec S + \oint_{\mathscr C^{\alpha \beta}} \vec \eta^{\alpha \beta} \cdot d\vec \ell + \iint_{\mathscr T^\beta} \vec B^\beta \cdot d\vec S$, where $\mathscr C^{\alpha \beta} \subset \mathscr T \cap U^\alpha \cap U^\beta$ is a loop cutting $\mathscr T$ in two parts. $\vec B^\alpha$ and $\vec \eta^{\alpha \beta}$ are the fields defining the degree 3 Deligne cohomology class. $\vec B$ is the $B$-field appearing in string theory\cite{string} previously cited. An usual magnetic potential $\vec A$ can appear also in 2-form electrodynamics to treat open strings ($\vec B$ being coupled with the string body and $\vec A$ being coupled with the string ends).\\

Clearly, we see an increase of the complexity of the gauge theories in the sequence : electrostatics $\to$ electromagnetism $\to$ 2-form electrodynamics (string theory). The Deligne degree appearing as a degree of complexity. What is the situation for the point of view of the adiabatic picture of quantum dynamics? Firstly note that a free evolution $|\psi(t) \rangle = e^{-\ihbar^{-1} \lambda_a(x)t} |a,x\rangle$ with $|\psi(0)\rangle = |a,x\rangle$ ($H(x)|a,x\rangle = \lambda_a(x)|a,x\rangle$, $x$ being not moved), is associated with a degree 1 Deligne class $[\lambda_a]_{\mathcal D}$ (the energy origin being arbitrary, it constitutes a gauge choice), $\lambda_a$ playing the role of an electric potential. The magnetic analogy shows clearly that the geometric phase of strong adiabatic transport defines a degree 2 Deligne class $[\chi,\vec A]_{\mathcal D}$ (where $\chi$ is the transition functions associated with magnetic monopoles - the eigenvalue crossings -). And finally the geometric phase of weak adiabatic transport defines a degree 3 Deligne class $[\xi,\vec \eta,\vec B]$ where $\vec B$ is the $B$-field. For the sake of simplicity, we have not introduce in the previous sections the fields $\xi$ and $\vec \eta$, but we can find explicit examples of these ones in ref.\cite{Viennot8} in the case of the Floquet adiabatic transport (a method to treat light-matter interaction, where we consider Floquet quasi-energy states which are entangled states between the matter states and the photon states). Another example can be found in ref.\cite{Viennot9} for the adiabatic transport of a quantum resonance where the interaction with the environment is not modelled by entanglement but by dissipation as in the example of the quantum wormhole with the single sheet model (as for eigenvalues crossing of the quantum wormhole, the resonance crossing is not a single point but a circle). Finally, we see table \ref{complx} the correspondance between the complexity of the dynamical system and the complexity of the geometric phase gauge theory.
\begin{table}
  \caption{\label{complx} Correspondance between quantum dynamical systems of control parameters $x$ and geometric phase gauge theory.}
  
  \begin{tabular}{l|c|l}
    \textbf{Quantum} & \textbf{Deligne} & \textbf{Electromagnetic} \\
    \textbf{dynamics} & \textbf{degree} & \textbf{analogy} \\
    \hline
    $x$ static & 1 & electrostatics \\
    \hline
    $x$ slowly moved & 2 & electromagnetism \\
    \hline
    $x$ slowly moved  & 3 & 2-form electrodynamics \\
    + entanglement & & (string theory) \\
    or dissipation & &
  \end{tabular}
\end{table}
This symmetry is intriguing from the point of view of the hierarchy of the physical theories. On the one hand, degree 3 gauge theory seems to be more fundamental than degree 2 gauge theory, since string theory corresponds to physics at a smaller size and higher energy than usual electrodynamics; but on the other hand, degree 3 gauge theory seems to be less fundamental than degree 2 gauge theory, since adiabatic transport with entanglement is the composition of two ``physical structures'' (adiabatic transport and entanglement dynamics). Since the increase of the complexity can induce the emergence of a theory similar to the descent to the most elementary, the concept of a ``most fundamental theory'' (or of a ``theory of everything'') defended in the reductionist attitude seems to be hardly difficult to properly define. The concept of emergence of a never-ending tower of theories related to each level of complexity or to each level of size/energy domain seems to be more consistent.\\

Remark: it is known that string theory needs addition of six compact extra-dimensions to the spacetime (not taken into account in the previous discussions in this paper). An emergent compact extra dimension to $\mathscr M$ appears also in Floquet adiabatic transport\cite{Viennot8}. Moreover, by a similar principle, it is shown in ref.\cite{Viennot10} that in the BFSS matrix model, six compact extra-dimensions emerge from the initial four spacetime dimensions due to the effects of the quantum vacuum fluctuations (playing the role of an environment).

\section{Conclusion}
Geometric phases in adiabatic quantum dynamics provide a concrete geometric realisation of the many-worlds of the Everett's interpretation of quantum mechanics. The main characteristics of this interpretation (``superposition'' of worlds, ``interferences'' between compatible worlds, emergence of the many-worlds structure due to the observation, relation between the preferred basis and decoherence phenomenon), have geometric realisations in the adiabatic picture taking the form of gauge theories onto copies of a control manifold. This geometric picture of the many-worlds interpretation is compatible with the matrix models of quantum gravity. We have seen a symmetry between the increase of complexity in adiabatic quantum dynamics and the descent to the most fundamental theories in electrodynamics (including the unification with gravity at the last stage of string theory). If we consider the geometric picture of adiabatic dynamics as the concrete realisation of the many-worlds interpretation, this shows, in contradiction with some criticisms concerning this one, that it is intimately related to quantum gravity: interferences between worlds (needed to have quantum transitions and changes of the probabilities of occupation) are strongly related to the quantum wormhole concept; gauge theory characteristic of string theory (degree 3 Deligne classes) emerges if we take into account entanglement with an environment (and so decoherence phenomenon) which is needed to solve the preferred basis problem.\\
What can be deduced, philosophically speaking, from the symmetry of the emergence of gauge theories following the descent to the fundamental and by the increase of the complexity of the dynamics? What can be deduced from the intimate link between the geometric phase formalism and the emergence of gravity in quantum spacetimes? Fundamentally, we have no difference between the control manifold of the adiabatic transport and the classical flat spacetime in which the observer positions its test particle (two ones denoted by $\mathscr M$ in the previous sections). These two manifolds are just parameter spaces in the mind of the observer needed only to have a representation of the dynamics and to organise the results of measurements by the quantum observables ($H(x)$, $X^i$, $\slashed D_x$). We have called ``analogy'' the symmetry between the two situations, geometric phases and electromagnetism/gravity. But the example of quantum gravity matrix model shows that it is maybe more than an analogy. We could consider that the mathematical structure (here a class of algebraic topology and its geometric realisation) defines the physical meaning. This point of view is sometimes invoked in the consideration of the ontological problem of quantum mechanics. At a middle point between the traditional object realist position (there is a Reality independent of the mind, accessible and intelligible, composed by objects) and the idealist position (the thing-in-itself is essentially unknowable in nature, we only put in order partial and biased representations of a Reality) or the instrumentalist position, we have structural realist positions. In these ones, we consider that the relations between the objects or the structures in which these objects take place, is the true essence of the Reality and not the objects themselves. It is possible to see the many-worlds interpretation as a structural realism since the many-worlds structure is relative to the relation system-observer. Another possibility consists to claim that familiar concepts from our mental images fail to render reality on the microscopic scale, only terms from abstract mathematics can describe it. So, in this position, mathematical structures are closest to the essences of the Reality than objects (particles, atoms,...). This position is called mathematical realism of physics, einsteinian realism or pythagorean realism\cite{Espagnat} (I prefer this last name, the first one induces confusion with platonicism which concerns the independent reality of mathematics, and the second one attributes to Einstein an adherence to this point of view, which is not obvious). In this pythagorean realist point of view, the emergence of quantum gravity (with curvature and torsion) and the emergence of many-worlds (with magnetic potentials and $B$-fields) in adiabatic dynamics with entanglement, are the same ``physical'' process consisting of the increase of the Deligne degree of the gauge theory (which is the only essence of the reality as the underlying mathematical structure). The geometric phases formalism could then be a link between the many-worlds interpretation and the pythagorean position in a syncretism of these two structural realisms. 

\appendix
\section{System-observer entangled state formulation}\label{RSF}
In the original formulation of his interpretation\cite{Everett}, Everett uses explicit relative states as states of a bipartite system constituted by the observed quantum system and the observer. We see here that it is possible to realise the same thing by a different way with the geometric phase formalism by using the Schr\"odinger-Koopman method\cite{Viennot11}. We consider anew the Schr\"odinger equation
\begin{equation}\label{SEq}
  \ihbar \partial_t |\psi(t)\rangle = H(x(t)) |\psi(t)\rangle
\end{equation}
with $|\psi\rangle \in \mathscr H$, $x \in \mathscr M$ and $H: \mathscr M \to \mathcal L (\mathscr H)$. In this equation, the observer appears only via the control parameters $x$ belonging to the control manifold $\mathscr M$. In the previous sections, we have considered that the evolution $t \mapsto x(t)$ is given a priori and forces the quantum system. Consider that this evolution results from a classical dynamics:
\begin{equation}\label{HamiltonEq}
  \dot x^i = \{x^i,\mathcal H\}
\end{equation}
where $\{\cdot,\cdot\}$ is the Poisson bracket of the classical system (observer) and $\mathcal H$ is its classical Hamiltonian. $x$ can be a set of couples degrees of freedom and conjugate momenta ($\mathscr M$ being then the phase space of the observer), eq.(\ref{HamiltonEq}) is then the Hamilton equations; or $x$ can be classical observables concerning the observer and eq.(\ref{HamiltonEq}) is the evolution equation of these ones. Note that this discussion can be generalised to non-Hamiltonian evolution (dissipative system)\cite{Viennot11} but for the sake of simplicity we do not consider here this possibility. Let the following equation be called the Schr\"odinger-Koopman equation:
\begin{equation}\label{SKEq}
  \ihbar \partial_t|\Psi(t)\rrangle = (H-\ihbar \mathscr L_{\mathcal H}) |\Psi(t)\rrangle
\end{equation}
where $|\Psi \rrangle \in \mathscr H \otimes L^2(\mathscr M,d\mu(x))$ ($\mu$ is a measure on $\mathscr M$ preserved by the Hamiltonian flow), $H(x)$ being considered as an operator of $\mathscr H \otimes L^2(\mathscr M,d\mu(x))$ (acting by multiplication in $L^2(\mathscr M,d\mu(x))$) and $\mathscr L_{\mathcal H} \in \mathcal L(L^2(\mathscr M,d\mu(x)))$ is defined by
\begin{equation}
  \forall f \in L^2(\mathscr M, d\mu(x)), \quad \mathscr L_{\mathcal H}(f) = \{f,\mathcal H\}
\end{equation}
We can prove (see ref.\cite{Viennot11}) that the solutions of eq.(\ref{SEq}) and eq.(\ref{SKEq}) are related by
\begin{equation}
  |\psi(t)\rangle = \langle x(t)|\Psi(t)\rrangle
\end{equation}
where $t \mapsto x(t)$ is solution of eq.(\ref{HamiltonEq}), $\langle x|$ being the linear functional providing the value at $x \in \mathscr M$ of the states of $L^2(\mathscr M,d\mu(x))$ ($\forall f \in L^2(\mathscr M,d\mu(x))$, $\langle x|f\rangle = f(x) \in \mathbb C$). By this fact, the Schr\"odinger-Koopman (SK) state $|\Psi\rrangle$ is another representation of the dynamics, but this one represents both the quantum system and its observer/control. It can then play the role of the relative state.\\
In the adiabatic limit without eigenvalue crossings we have
\begin{equation}
  \langle x(t)|\Psi(t)\rrangle \simeq \sum_a c_a e^{-\imath \varphi_a(t)} |a,x(t)\rrangle
\end{equation}
with $\varphi_a(t) = \hbar^{-1} \int_0^t \lambda_a(x(t'))dt' + \int_0^t A^a_i(x(t'))\dot x^i(t')dt'$ ($\lambda_a$ and $|a,x\rangle$ being the eigenvalues and eigenvectors of $H$ and $\vec A^a = -\imath \langle a,x|\vec \nabla |a,x\rangle$). A possible entangled state can be then (in the adiabatic approximation)
\begin{equation}
  |\Psi(t)\rrangle \simeq \int \sum_a c_a e^{-\imath \varphi_a(t)} |a,x(t)\rangle \otimes |x(t)\rangle \mathscr D[x(t)]
\end{equation}
where $\int \mathscr D[x(t)]$ denotes the Feynman path integral (restricted onto paths solutions of eq.(\ref{HamiltonEq})). $\varphi_a$ plays the role of the classical action for the world/branche $\mathscr M_a$ in accordance with the electromagnetic analogy ($\mathcal L_a = \hbar^{-1} \lambda_a + \vec A^a \cdot \frac{d\vec x}{dt}$ being the Lagrangian of the interaction of a particle of unit charge with an electromagnetic field of scalar potential $-\hbar^{-1} \lambda_a$ and vector potential $\vec A$).\\
$|\Psi(t)\rrangle$ is the state of the bipartite system constituted by the quantum system and the observer (in the meaning of the experiment device used by the observer to control the quantum system). $|x \rangle$ is the observer state where the control parameters have the values $x$. $\int e^{-\imath \varphi_a(t)} |a,x(t)\rangle \otimes |x(t)\rangle \mathscr D[x(t)]$ can be interpreted as the state of the bipartite system where a measurement of $H$ has been realised and has provided the result $\lambda_a$ independently from the evolution $t \mapsto x(t)$ followed by the observer (since we have a quantum superposition of all possible evolutions). And then, in $|\Psi(t)\rrangle$, this part represents the state of the world with the outcome of the measurement of $H$ is $\lambda_a$ independently from the history of the observer. In this meaning, $|\Psi \rrangle$ is the state of the many worlds structure including all possibilities for the system and the observer (the ``state of the Universe''). The observer ``viewing'' the outcome $\lambda_a$ after measurement ``lives'' in the world of state $\int c_a e^{-\imath \varphi_a(t)} |a,x(t)\rangle \otimes |x(t)\rangle \mathscr D[x(t)]$ (relative to this outcome). In contrast with the Copenhagen interpretation, the measurement does not induce a wave packet reduction, the state of the ``Universe'' $|\Psi\rrangle$ is not modified, the measurement is just the ``disclosure'' of the world in which the observer ``lives'' relatively to the observed outcome. $|\Psi \rrangle$ is the state of the many worlds structure in the meaning of it is the state independent of a particular outcome of the measurement and independent of the history of the observer. We can see an emerging structure depending on the choice of the considered measurement by considering a basis change. Let $O$ be another observable, $(|i\rangle)_i$ be its eigenbasis and $(\mu_i)_i$ be its spectrum. In this new basis we have
\begin{equation}
  |\Psi(t)\rrangle \simeq \int \sum_i k_i(x(t)) |i\rangle \otimes |x(t) \rangle \mathscr D[x(t)]
\end{equation}
with $k_i(x(t)) = \sum_a c_a e^{-\imath \varphi_a(t)} \langle i|a,x(t)\rangle$. In a same way, $\int k_i(x(t)) |i\rangle \otimes |x(t) \rangle \mathscr D[x(t)]$ is the state of the world relative to the outcome $\mu_i$ of the measurement of $O$. Note that this approach is different from the original Everett's approach of the state of the Universe which does not involve path integral formulation. The introduction of this formulation in the present context is due to the use of the adiabatic approximation which implies that the evolution of the observer is classical. The path integral formulation with the Schr\"odinger-Koopman formalism are used to treat the entanglement in a bipartite system mixing quantum and classical degrees of freedom. In the Everett's theory, no quantum-classical mixture occurs, and all objects of the universe are considered quantum. 

\section*{Data availability statements}
No data is used in this work.

\section*{Conflict of interest}
On behalf of all authors, the corresponding author states that there is no conflict of interest.


\begin{thebibliography}{0}
\bibitem{Bitbol} M. Bitbol, {\it Physique et philosophie de l'esprit} (Champs-Flammarion, Paris, 2000).
\bibitem{Omnes} R. Omn\`es, {\it Quantum philosophy: understanding and interpreting contemporary science} (Princeton University Press, Princeton, 2002).
\bibitem{Espagnat} B. d'Espagnat, {\it On physics and philosophy} (Princeton University Press, Princeton, 2006).
\bibitem{Everett} H. Everett, {\it Rev. Mod. Phys.} {\bf 29}, 454 (1957).
\bibitem{Berry} M.V. Berry, {\it Proc. R. Soc. Lond. A} {\bf 392}, 45 (1984).
\bibitem{Shapere} A. Shapere and F. Wilczek, {\it Geometric phases in physics} (World Scientific, Singapore, 1989).
\bibitem{Simon} B. Simon, {\it Phys. Rev. Lett.} {\bf 51}, 2167 (1983).
\bibitem{Moody} J. Moody, A. Shapere and F. Wilczek, {\it Phys. Rev. Lett.} {\bf 56}, 893 (1086).
\bibitem{Teufel} S. Teufel, {\it Adiabatic perturbation theory in quantum dynamics} (Springer, Berlin, 2003).
\bibitem{Aharonov} Y. Aharonov and D. Bohm, {\it Phys. Rev.} {\bf 115}, 485 (1959).
\bibitem{Wilczek} F. Wilczek and A. Zee, {\it Phys. Rev. Lett.} {\bf 52}, 2111 (1984).
\bibitem{Aharonov2} Y. Aharonov and J. Anandan, {\it  Phys. Rev. Lett.} {\bf 58}, 1593 (1987).
\bibitem{Nenciu} G. Nenciu, {\it J. Phys. A: Math. Gen.} {\bf 13} L15 (1980).
\bibitem{Viennot1} D. Viennot, {\it J. Math. Phys.} {\bf 46} 072102 (2005).
\bibitem{Frankel} chapter 16.4e in T. Frankel, {\it The geometry of physics} (Cambridge University Press, Cambridge, 1997).
\bibitem{Boscain} U. Boscain, F.C. Chittaro, P. Mason and M. Sigalotti, {\it IEEE Transactions on automatic control} {\bf 57} 1970 (2012).
\bibitem{Wallas} S. Saunders, J. Barrett, A. Kent and D. Wallas, {\it Many Worlds? Everett, quantum theory, and reality} (Oxford University Press, Oxford, 2010).
\bibitem{Breuer} H.P. Breuer and F. Petruccione, {\it The theory of open quantum systems} (Oxford University Press, Oxford, 2002).
\bibitem{Viennot2} D. Viennot and L. Aubourg, {\it J. Phys. A: Math. Theor.} {\bf 48} 025301 (2015).
\bibitem{Viennot3} D. Viennot, {\it J. Geom. Phys.} {\bf 133} 42 (2018).
\bibitem{Viennot4} D. Viennot and J. Lages, {\it J. Phys. A: Math. Theor.} {\bf 44} 365301 (2011).
\bibitem{Viennot5} D. Viennot, {\it J. Phys. A: Math. Theor.} {\bf 47} 295301 (2014).
\bibitem{string} J.-G. Zhou, {\it Nuclear Physics B} {\bf 607} 237 (2001).
\bibitem{Penrose} R. Penrose, {\it Science Watch}, august 1991.
\bibitem{string2} B. Zwiebach, {\it A first course in string theory} (Cambridge University Press, Cambridge, 2009).
\bibitem{QLG} C. Rovelli and F. Vidotto, {\it Covariant loop quantum gravity} (Cambridge University Press, Cambridge, 2015).
\bibitem{NCG} M. Buric and J. Madore, arXiv:hep-th/0406232 (2005).
\bibitem{EmG} S. De, T.P. Singh and A. Varma, {\it Int. J. Modern Phys. D} {\bf 28} 1944003 (2019).
\bibitem{Schneiderbauer} L. Schneiderbauer and H.C. Steinacker {\it J. Phys. A: Math. Theor.} {\bf 49}, 285301 (2016).
\bibitem{fuzzy} A. Sykora arXiv:1610.01504 (2016).
\bibitem{BFSS} T. Banks, W. Fischler, S.H. Shenker and L. Susskind, {\it Phys. Rev. D} {\bf 55}, 5112 (1997).
\bibitem{Klammer} D. Klammer and H. Steinacker, {\it J. High Energy Phys.} {\bf JHEP08(2008)}; 074 (2008).
\bibitem{Steinacker} H. Steinacker {\it Class. Quantum Grav.} {\bf 27}, 133001 (2010).
\bibitem{Kunter} N. Kunter and H. Steinacker {\it J. Geom. Phys.} {\bf 62}, 1760 (2012).
\bibitem{Perelomov} A. Perelomov {\it Generalized Coherent States and Their Applications} (Springer, Berlin, 1986).
\bibitem{Viennot6} D. Viennot, {\it Class. Quantum Grav.} {\bf 38}, 245004 (2021).
\bibitem{Wallace2} D. Wallace, {\it Studies in History and Philosophy of Modern Physics} {\bf 33}, 637 (2002).
\bibitem{Steinacker2} H.C. Steinacker {\it J. High Energy Phys.} {\bf JHEP04(2020)}, 111 (2020).
\bibitem{Einstein} A. Einstein and N. Rosen {\it Phys. Rev.} {\bf 48}, 73 (1935).
\bibitem{Flamm} L. Flamm {\it Physikalische Zeitschrift} {\bf XVII}, 448 (1916).
\bibitem{Viennot7} D. Viennot, {\it J. Math. Phys.} {\bf 63}, 082302 (2022).
\bibitem{Morris} M.S. Morris and K.S. Thorne, {\it American Journal of Physics} {\bf 56}, 395 (1988).
\bibitem{Anderson} P.W. Anderson, {\it Science} {\bf 177}, 393 (1972).
\bibitem{Langton} C.G. Langton, {\it Physica D: Nonlinear Phenomena} {\bf 22}, 120 (1986).
\bibitem{Brylinski} J.-L. Bylinski, {\it Loop spaces, characteristic classes and geometric quantisation} (Birkh\"auser, Boston, 1993).
\bibitem{Henneaux} M. Henneaux and C. Teitelboim, {\it Foundations of Physics} {\bf 16}, 593 (1986).
\bibitem{Viennot8} D. Viennot, {\it J. Phys. A: Math. Theor.} {\bf 42}, 395302 (2009).
\bibitem{Viennot9} D. Viennot, {\it J. Math. Phys.} {\bf 50}, 052101 (2009).
\bibitem{Viennot10} D. Viennot and L. Aubourg, {\it Class. Quantum Grav.} {\bf 35}, 135007 (2018).
\bibitem{Viennot11} D. Viennot and L. Aubourg, {\it J. Phys. A: Math. Theor.} {\bf 51}, 335201 (2018).
\end{thebibliography}
\end{document}